\documentclass[final,3p,times]{elsarticle}
\usepackage{amsfonts}

\usepackage{amssymb}
\usepackage{amsmath}
\usepackage{graphics}
\usepackage{epsfig}
\usepackage{epstopdf}
\usepackage{amsthm}
\usepackage{textcomp}
\usepackage{color}
\usepackage{float}
\usepackage{multirow}
\biboptions{sort&compress}

\journal{Nonlinear Analysis: Real World Applications}

\begin{document}

\begin{frontmatter}

\title{Analysis of the effectiveness of the truth-spreading strategy for inhibiting rumors}

\cortext[cor1]{Corresponding author}
%%\cortext[cor2]{Principal corresponding author}

\author[label1]{Lu-Xing Yang}
\ead{ylx910920@gmail.com}

\author[label2]{Pengdeng Li}
\ead{1414797521@qq.com}

\author[label2]{Xiaofan Yang\corref{cor1}}
\ead{xfyang1964@gmail.com}

\author[label2]{Yingbo Wu}
\ead{wyb@cqu.edu.cn}

\author[label3]{Yuan Yan Tang}
\ead{yytang@umac.mo}

\address[label1]{Faculty of Electrical Engineering, Mathematics and Computer Science, Delft University of Technology, Delft, GA 2600, The Netherlands}

\address[label2]{School of Software Engineering, Chongqing University, Chongqing, 400044, China}

\address[label3]{Department of Computer and Infomation Science, The University of Macau, Macau}

\begin{abstract}
Spreading truths is recognized as a feasible strategy for inhibiting rumors. This paper is devoted to assessing the effectiveness of the truth-spreading strategy. An individual-level rumor-truth spreading model (the generic URTU model) is derived. Under the model, two criteria for the termination of a rumor are presented. These criteria capture the influence of the network structures on the effectiveness of the truth-spreading strategy. Extensive simulations show that, when the rumor or the truth terminates, the dynamics of a simplified URTU model (the linear URTU model) fits well with the actual rumor-truth interplay process. Therefore, the generic URTU model forms a theoretical basis for assessing the effectiveness of the truth-spreading strategy for restraining rumors.
\end{abstract}

\begin{keyword}
rumor/truth spreading \sep rumor/truth spreading network \sep generic/linear rumor-truth spreading model \sep spectral radius

\end{keyword}

\end{frontmatter}

% \linenumbers


\section{Introduction}

Rumors are loosely defined as unconfirmed elaborations or annotations of public things, events or issues. As an important form of social interactions, rumor spreading has a significant impact on human affairs. The rapidly popularized online social networks (OSNs) offer a shortcut for the fast spread of rumors, greatly enlarging their influence \cite{Viswanath2009, Kwak2010, Doerr2012}. Unfortunately, most rumors could induce social panic or economic loss \cite{Thomas2007}. For example, Syrian hackers once broke into the twitter account of Associated Press (AP) and dispersed the rumor that explosions at White House had injured Obama, leading to 10 billion USD losses before the rumor was clarified \cite{Peter2013}. Therefore, one of the major concerns in the field of cybersecurity is to contain the prevalence of rumors in OSNs \cite{Budak2011}.

The rumor spreading dynamics is intended to model and study the spreading process of rumors, so as to gain insight into the influence of different factors on the prevalence of rumors and thereby to work out cost-effective strategies of restraining rumors. In 1964, Daley and Kendall \cite{Daley1964} introduced the first rumor spreading model (the DK model) by modeling the spreading process of rumors as a discrete-time iterative process on fully connected networks. Later, Maki and Thompson \cite{Maki1973} advised a variant of the DK model (the MT model). In his PhD thesis, Cintron-Arias \cite{Cintron2006} suggested a differential dynamical system model capturing the spread of rumors by introducing the state transition rates to the DK model. Since then, a multitude of rumor spreading models on homogeneous networks have been proposed, aiming at understanding the influence of different factors, such as the memory \cite{ZhaoLJ2011a, ZhaoLJ2012a}, the skepticism and denial \cite{HuangWT2011}, the incubation \cite{HuoLA2012a, ZhaoLJ2015}, the education or scientific knowledge \cite{Afassinou2014, HuoLA2016}, the latency \cite{HuoLA2015}, the government behavior \cite{XuJP2016} and others \cite{HuoLA2017}, on the prevalence of rumors.

A series of emperical studies starting at the end of last century show that, surprisingly, OSNs are heterogeneous rather than homogeneous \cite{Albert2002, Ebel2002}. From then on, much effort has been put to inspect the spread of rumors on complex networks with the aid of the mean-field theory, with emphasis on the influence of different factors such as the network heterogeneity \cite{ZhouJ2007}, the nonlinear spreading rate \cite{Roshani2012}, the incubation \cite{Singh2012}, the memory \cite{ZhaoLJ2013a, ZhaoLJ2013b}, the trust between individuals \cite{WangYQ2013}, the countermeasures \cite{ZanYL2014, HeZB2015, HeZB2017}, the latency \cite{XiaLL2015, LiuQM2017}, the time-varying parameters \cite{QiuXY2016}, the government punishment \cite{LiD2017}, the heterogeneous transmission \cite{Oliveros2017} and others \cite{Nekovee2007, Naimi2013}. As these models are quite rough, their dynamics may severely deviate from the actual rumor spreading process. Furthermore, the genuine influence of the network topology on the rumor spreading cannot be made certain by studying these models.

In 2009, Van Mieghem et al. \cite{Mieghem2009} proposed a (continuous-time) individual-level epidemic model (the exact SIS model), which accurately captures the average dynamics of the SIS epidemic \cite{Mieghem2009, Mieghem2011}. In pursuit of this idea, a number of individual-level epidemic models have recently been reported \cite{Sahneh2012, Sahneh2013, YangLX2015, XuSH2015, ZhengR2015, YangLX2017a, YangLX2017b}. Continuous-time individual-level models are especially efficacious in exploring the impact of the network topology on the epidemic process and, hence, are suited to the study of rumor spreading. To our knowledge, so far there is no report in this aspect.

Clarifying rumors by spreading truths is a feasible strategy for inhibiting rumors in OSNs \cite{WenS2014, WenS2015}. Before the appearance of a rumor and the truth, all persons in the OSN are uncertain. After the appearance of the rumor and the truth, depending on personal judgement or/and knowledge, a person either believes the rumor, or believes the truth, or is uncertain (i.e., believe neither the rumor nor the truth). This paper is devoted to assessing the effectiveness of the truth-spreading strategy. An individual-level rumor-truth spreading model (the generic URTU model) is derived. Then two criteria for the extinction of a rumor are presented, which capture the influence of the network structures on the effectiveness of the truth-spreading strategy. Extensive simulations show that, when the rumor or the truth tends to extinction, the dynamics of a simplified URTU model (the linear URTU model) fits well with the actual rumor-truth interplay process. Therefore, the generic URTU model provides a proper basis for assessing the effectiveness of the truth-spreading strategy for restraining rumors.

The subsequent materials are organized in this fashion. Section 2 derives the generic URTU model as well as the linear URTU model. Section 3 studies the dynamics of the generic URTU model. Experimental results examining the effectiveness of the linear URTU model are reported in Section 4. Finally, Section 5 summarizes this work.

\newtheorem{rk}{Remark}
\newproof{pf}{Proof}
\newtheorem{thm}{Theorem}
\newtheorem{lm}{Lemma}
\newtheorem{exm}{Example}
\newtheorem{cor}{Corollary}
\newtheorem{de}{Definition}
\newtheorem{cl}{Claim}
\newtheorem{pro}{Proposition}
\newtheorem{con}{Conjecture}

\newproof{pfcl1}{Proof of Claim 1}
\newproof{pfcl2}{Proof of Claim 2}

\section{A generic rumor-truth spreading model}

This section is dedicated to establishing a generic continuous-time dynamic model capturing the interaction between a rumor and the truth.

\subsection{Notions, notations and hypotheses}

Suppose a rumor and the truth spread in an OSN consisting of $N$ persons labelled $1, 2, ..., N$. Let $V = \{1, 2, \cdots, N\}$. Suppose the rumor is propagated through a rumor-spreading network $G_R = (V, E_R)$, where $(i, j) \in E_R$ if and only if person $j$ can tell the rumor to person $i$. Suppose the truth is circulated through a truth-spreading network $G_T = (V, E_T)$, where $(i, j) \in E_T$ if and only if person $j$ can tell the truth to person $i$. In what follows, the two networks are always assumed to be strongly connected.

At the beginning, there is neither the rumor nor the truth, so all persons in the OSN are uncertain. After the appearance of the rumor and the truth, every person in the OSN is assumed to be in one of three possible states: \emph{rumor-believing}, \emph{truth-believing}, and \emph{uncertain}. A rumor-believing person believes the rumor, a truth-believing person believes the truth, and an uncertain person believes neither the rumor nor the truth. Depending on personal judgement on the event, every person may choose to believe the rumor, or to believe the truth, or to be uncertain. Let $X_i(t)$ = 0, 1, and 2 denote that, at time $t$, person $i$ is uncertain, rumor-believing, and truth-believing, respectively. Then the state of the OSN at time $t$ is represented by the vector
\[
\mathbf{X}(t) = (X_1(t), X_2(t), \cdots, X_N(t))^T.
\]

Next, let us introduce a set of hypotheses as follows.

\begin{enumerate}
	
	\item[(H$_1$)] Due to the influence of a rumor-believer $j$, at any time an uncertain person $i$ turns to believe the rumor at rate $\beta_{ij}^U \geq 0$. Here, $\beta_{ij}^U > 0$ if and only if $(i, j) \in E_R$. $\beta_{ij}^U$ is proportional to (a) the rate at which person $j$ tells the rumor to person $i$, and (b) the probability of person $i$ believing the rumor when hearing it. Let $\mathbf{B}_U = \left(\beta_{ij}^U\right)_{N \times N}$. This hypothesis captures the influence of the rumor spreading on uncertain persons.
	
	\item[(H$_2$)] Due to the influence of a rumor-believer $j$, at any time a truth-believer $i$ turns to believe the rumor at rate $\beta_{ij}^T \geq 0$. Here, $\beta_{ij}^T > 0$ if and only if $(i, j) \in E_R$. $\beta_{ij}^T$ is proportional to (a) the rate at which person $j$ tells the rumor to person $i$, and (b) the probability of person $i$ believing the rumor when hearing it. Let $\mathbf{B}_T = \left(\beta_{ij}^T\right)_{N \times N}$. This hypothesis captures the influence of the rumor spreading on truth-believers. Certainly, we have $\beta_{ij}^T \leq \beta_{ij}^U$.
	
	\item[(H$_3$)] Due to the influence of a truth-believer $j$, at any time an uncertain person $i$ turns to believe the truth at rate $\gamma_{ij}^U \geq 0$, Here, $\gamma_{ij}^U > 0$ if and only if $(i, j) \in E_T$. $\gamma_{ij}^U$ is proportional to (a) the rate at which person $j$ tells the truth to person $i$, and (b) the probability of person $i$ believing the truth when hearing it. Let $\mathbf{C}_U = \left(\gamma_{ij}^U\right)_{N \times N}$. This hypothesis captures the influence of the truth spreading on uncertain persons.

	\item[(H$_4$)] Due to the influence of a truth-believer $j$, at any time a rumor-believer $i$ turns to believe the truth at rate $\gamma_{ij}^R \geq 0$, Here, $\gamma_{ij}^R > 0$ if and only if $(i, j) \in E_T$. $\gamma_{ij}^R$ is proportional to (a) the rate at which person $j$ tells the truth to person $i$, and (b) the probability of person $i$ believing the truth when hearing it. Let $\mathbf{C}_R = \left(\gamma_{ij}^R\right)_{N \times N}$. This hypothesis captures the influence of the truth spreading on rumor-believers. Certainly, we have $\gamma_{ij}^R \leq \gamma_{ij}^U$.
	
	\item[(H$_5$)] Due to the forgetfulness or the loss of interest, a rumor-believer $i$ turns to be uncertain at rate $\delta_i^R>0$. Let $\mathbf{D}_R = diag\left(\delta_i^R\right)$.
	
	\item[(H$_6$)] Due to the forgetfulness or the loss of interest, a truth-believer $i$ turns to be uncertain at rate $\delta_i^T>0$. Let $\mathbf{D}_T = diag\left(\delta_i^T\right)$.
	
\end{enumerate}

All the forthcoming rumor-truth spreading models are assumed to comply with these hypotheses.

\subsection{The original URTU model}

For fundamental knowledge on continuous-time Markov chain, see Ref. \cite{Stewart2009}.

Another way of representing the group state at time $t$ is by the decimal number $i(t)=\sum_{k=1}^{N}X_k(t)3^{k-1}$. In this context, there are totally $3^N$ possible OSN states: $0, 1, \cdots, 3^{N} - 1$. According to the previous hypotheses, the infinitesimal generator $\mathbf{Q}=\left[q_{ij}\right]_{3^N \times 3^N}$ for the rumor-truth spreading process is given as

\begin{small}
	\begin{equation}
	q_{ij}=\left\{
	\begin{aligned}
	&\delta_m^R   & \text{if} &\quad i=j+3^{m-1}; \\
	& & & \quad m=1,2,\cdots,N, x_{m}=1 \\
	&\delta_m^T  & \text{if} &\quad i=j+2\cdot3^{m-1}; \\
	& & & \quad m=1,2,\cdots,N, x_{m}=2 \\
	&\sum_{k=1}^{N}\beta_{mk}^U1_{\{x_{k}=1\}}   & \text{if} &\quad i=j-3^{m-1}; \\
	& & & \quad m=1,2,\cdots,N, x_{m}=0  \\
	&\sum_{k=1}^{N}\beta_{mk}^T1_{\{x_{k}=1\}}   & \text{if} &\quad i=j+3^{m-1}; \\
	& & & \quad m=1,2,\cdots,N, x_{m}=2  \\
	&\sum_{k=1}^{N}\gamma_{mk}^U1_{\{x_{k}=2\}}  & \text{if}&
	\quad i=j-2\cdot3^{m-1}; \\
	& & & \quad m=1,2,\cdots,N, x_{m}=0 \\
	&\sum_{k=1}^{N}\gamma_{mk}^R1_{\{x_{k}=2\}}  & \text{if}&
	\quad i=j-3^{m-1}; \\
	& & & \quad m=1,2,\cdots,N, x_{m}=1 \\
	&-\sum_{k=0,k\neq i}^{N-1}q_{ik}  & \text{if} & \quad i=j\\
	& 0 \quad  & \quad  &\text{otherwise}.
	\end{aligned}
	\right.
	\end{equation}
\end{small}

\noindent where $i=\sum_{k=1}^{N}x_{k}3^{k-1}$, $1_A$ stands for the indicator function of set $A$. The continuous-time Markov chain model with the infinitesimal generator $\mathbf{Q}$ is referred to as the \emph{original Uncertain-Rumor-Truth-Uncertain (URTU) model}. See Fig. 1 for the state transition rates of a person under the original URTU model.

\begin{figure}[!t]
	\centering
	\includegraphics[width=2.5in]{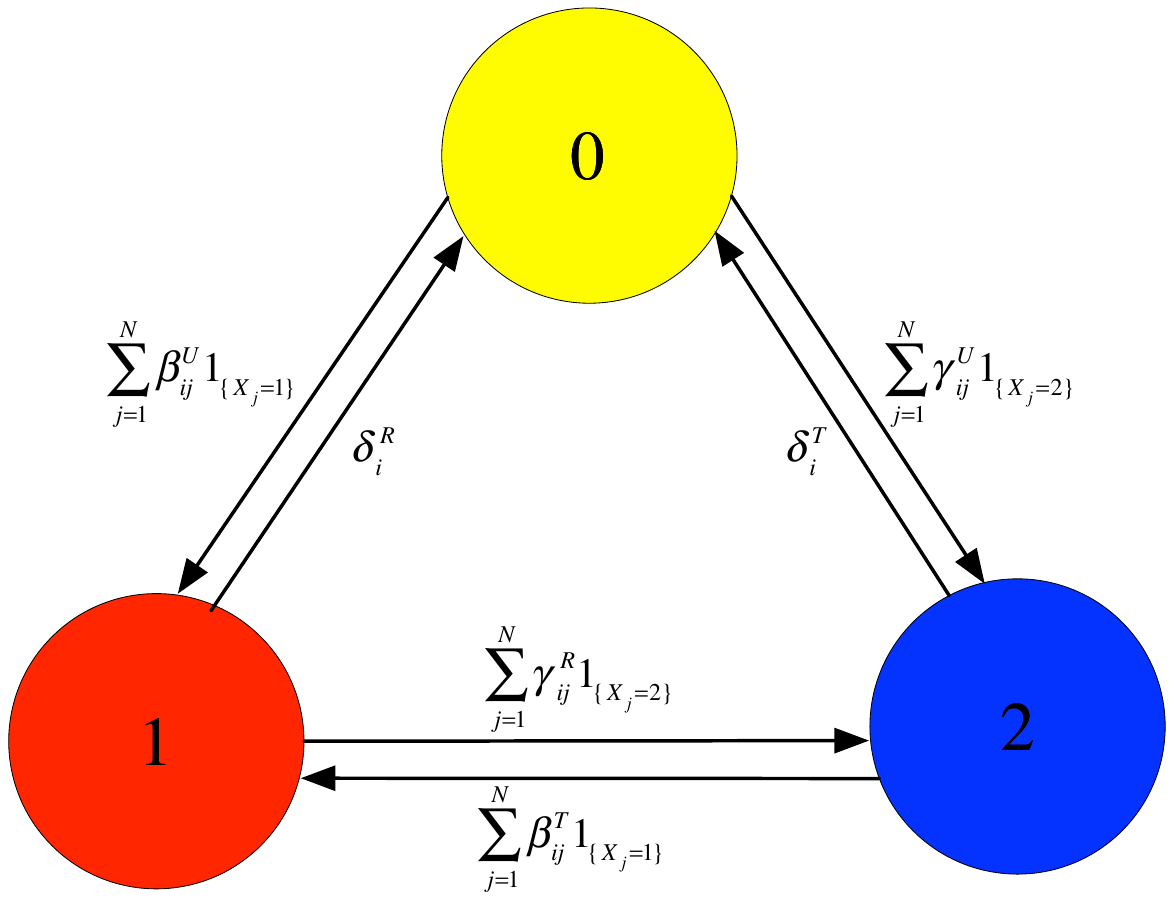}
	\caption{The state transition rates of person $i$ under the original URTU model.}
\end{figure}

\subsection{The exact URTU model}

Let $s_i(t)$ denote the probability that the group state at time $t$ is $i = \sum_{k=1}^{N}x_{k}3^{k-1}$. That is,
\[
s_{i}(t)=\Pr\left\{X_{1}(t)=x_{1},\cdots,X_{N}(t)=x_N\right\}.
\]
Let $\mathbf{s}(t)=[s_{0}(t),\cdots,s_{3^{N}-1}(t)]^T$. Then $\mathbf{s}(t)$ obeys
\begin{equation}
\frac{d\mathbf{s}^{T}(t)}{dt}=\mathbf{s}^{T}(t)\mathbf{Q}.
\end{equation}
\noindent This continuous-time Markov chain model accurately captures the average dynamics of the rumor-truth interaction. Therefore, we refer to the model as the \emph{exact URTU model}. The state transition rates of a person under the exact URTU model cannot be clearly shown as a diagram.

Although the exact URTU model is a linear differential system, with the solution $\mathbf{s}^{T}(t)=\mathbf{s}^{T}(0)e^{\mathbf{Q}t}$, its dimensionality grows exponentially with the increasing size of the OSN, leading to mathematical intractability.

Let
\[
R_i(t) = \Pr\{X_i(t) = 1\}, T_i(t) = \Pr\{X_i(t) = 2\}.
\]
$R_i(t)$ and $T_i(t)$ probabilistically capture the state of person $i$ at time $t$. The following lemma gives an equivalent form of the exact URTU model.

\begin{lm}
	The exact URTU model is equivalent to the model
	\begin{small}
		\begin{equation}
		\left\{
		\begin{aligned}
		\frac{dR_i(t)}{dt} &= \sum_{j = 1}^N \beta_{ij}^U \Pr\{X_i(t) = 0, X_j(t) = 1\} +\sum_{j = 1}^N \beta_{ij}^T \Pr\{X_i(t) = 2, X_j(t) = 1\} -\sum_{j = 1}^N \gamma_{ij}^R \Pr\{X_i(t) = 1, X_j(t) = 2\} - \delta_i^R R_i(t), \\
		\frac{dT_i(t)}{dt} &= \sum_{j = 1}^N \gamma_{ij}^U \Pr\{X_i(t) = 0, X_j(t) = 2\} +\sum_{j = 1}^N \gamma_{ij}^R \Pr\{X_i(t) = 1, X_j(t) = 2\} -\sum_{j = 1}^N \beta_{ij}^T \Pr\{X_i(t) = 2, X_j(t) = 1\} - \delta_i^T T_i(t), \\
		& \quad \quad i = 1, 2, \cdots, N.
		\end{aligned}
		\right.
		\end{equation}
	\end{small}
\end{lm}

The proof of this lemma is left to Appendix A. The equivalent model is not closed. If one attempted to close the equivalent model by adding more joint probability terms, the resulting model would be of dimensionality $3^N$ again, which is still mathematically intractable.

\subsection{The linear URTU model}

In order to simplify the exact URTU model, it is necessary to reduce its dimensionality while keeping its closedness. To this end, let us make an added
set of hypotheses as follows. For $1 \leq i, j \leq N, i \neq j$.

\begin{enumerate}
	\item[(H$_7$)] $\Pr\{X_i(t) = 0, X_j(t) = 1\} = (1 - R_i(t) - T_i(t))R_j(t).$
	\item[(H$_8$)] $\Pr\{X_i(t) = 0, X_j(t) = 2\} = (1 - R_i(t) - T_i(t))T_j(t).$
	\item[(H$_9$)] $\Pr\{X_i(t) = 1, X_j(t) = 2\} = R_i(t) T_j(t)$.
	\item[(H$_{10}$)]  $\Pr\{X_i(t) = 2, X_j(t) = 1\} = T_i(t) R_j(t)$.
\end{enumerate}

\noindent These hypotheses are known as the independence hypotheses. Based on the equivalent model (3) and these four hypotheses, we obtain the following approximation model of the exact URTU model.
\begin{small}
	\begin{equation}
	\left\{
	\begin{aligned}
	\frac{dR_i(t)}{dt} &= (1-R_i(t)-T_i(t))\sum_{j = 1}^N \beta_{ij}^U R_j(t)- \delta_i^R R_i(t) +T_i(t)\sum_{j = 1}^N \beta_{ij}^T R_j(t)-R_i(t)\sum_{j = 1}^N \gamma_{ij}^R T_j(t) , \\
	\frac{dT_i(t)}{dt} &= (1-R_i(t)-T_i(t))\sum_{j = 1}^N \gamma_{ij}^U T_j(t)- \delta_i^T T_i(t) +R_i(t)\sum_{j = 1}^N \gamma_{ij}^R T_j(t)-T_i(t)\sum_{j = 1}^N \beta_{ij}^T R_j(t) , \\
	& \quad \quad i = 1, 2, \cdots, N.
	\end{aligned}
	\right.
	\end{equation}
\end{small}
\noindent We refer to this model as the \emph{linear URTU model}, because the rumor-spreading rates, $\sum_{j = 1}^N \beta_{ij}^U R_j(t)$ and $\sum_{j = 1}^N \beta_{ij}^T R_j(t)$, are linear in $R_1(t), \cdots, R_N(t)$, and the truth-spreading rates, $\sum_{j = 1}^N \gamma_{ij}^U T_j(t)$ and $\sum_{j = 1}^N \gamma_{ij}^R T_j(t)$, are linear in $T_1(t), \cdots, T_N(t)$. See Fig. 2 for the state transition rates of a person under the linear URTU model.

\begin{figure}[!t]
	\centering
	\includegraphics[width=2.5in]{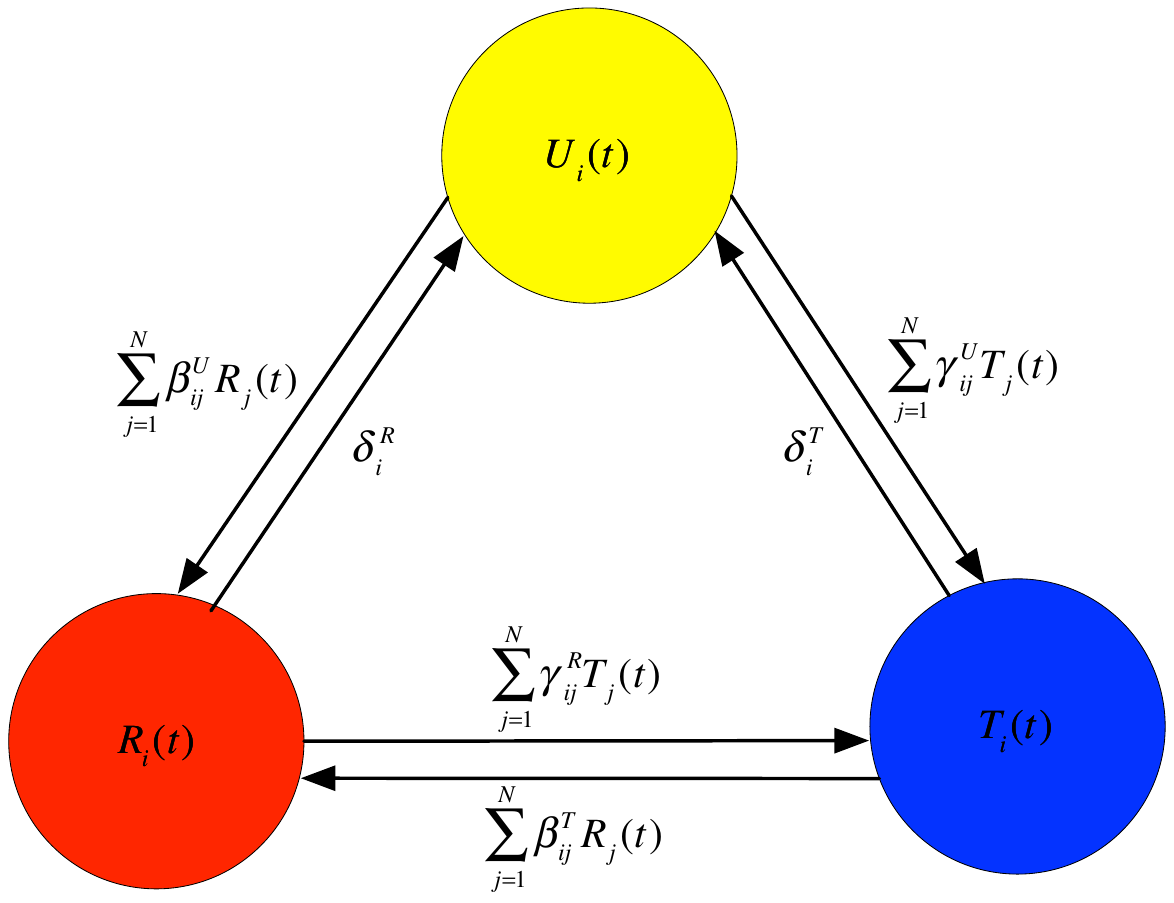}
	\caption{The state transition rates of person $i$ under the linear URTU model.}
\end{figure}

The linear URTU model is a closed 3$N$-dimensional dynamical system and is mathematically tractable. However, as the linear spreading rates may be different from the actual spreading rates, the dynamics of the model may deviate from the actual average dynamics of rumor-truth interaction.

\subsection{The generic URTU model}

For the purpose of approximating the exact URTU model more accurately, let us consider a more general rumor-truth spreading model as follows.
\begin{small}
	\begin{equation}
	\left\{
	\begin{aligned}
	\frac{dR_i(t)}{dt}&=\left(1-R_i(t)-T_i(t)\right)f_i^U\left(R_1(t), \cdots, R_N(t)\right)-\delta_i^R R_i(t) +T_i(t)f_i^T\left(R_1(t), \cdots, R_N(t)\right) -R_i(t)g_i^R(T_1(t),\cdots,T_N(t)),\\
	\frac{dT_i(t)}{dt}&=\left(1-R_i(t)-T_i(t)\right)g_i^U\left(T_1(t), \cdots, T_N(t)\right)-\delta_i^TT_i(t) +R_i(t)g_i^R\left(T_1(t), \cdots, T_N(t)\right) -T_i(t)f_i^T(R_1(t),\cdots,R_N(t)), \\
	&\quad \quad i=1,2,\cdots, N.
	\end{aligned}
	\right.
	\end{equation}
\end{small}
\noindent Here, the spreading rates are assumed to satisfy the following generic conditions.

\begin{enumerate}
	
	\item[(C$_1$)] (Proximity) An uncertain person or a truth-believer can and can only be influenced by those rumor-believers that can send the rumor to it through the rumor-spreading network. That is, $f_i^U$ and $f_i^T$ is dependent upon $x_j$ if and only if $(i,j) \in E_R$. Likewise, An uncertain person or a rumor-believer can and can only be influenced by those truth-believers that can send the truth to it through the truth-spreading network. That is, $g_i^U$ and $g_i^R$ is dependent upon $x_j$ if and only if $(i,j) \in E_T$.
	
	\item[(C$_2$)] (Nullity) The rumor cannot spread unless there is a rumor-believer in the group. That is, $f_i^U(0, \cdots, 0)=f_i^T(0, \cdots, 0)=0$. Likewise, the truth cannot spread unless there is a truth-believer in the group. That is, $g_i^U(0, \cdots, 0)=g_i^R(0, \cdots, 0)=0$.
	
	\item[(C$_3$)] (Ordering) As compared to a truth-believer, an uncertain person is easier to believe the rumor. As compared to a rumor-believer, an uncertain person is easier to believe the truth. That is, $f_i^U \geq f_i^T$, $g_i^U\geq g_i^R$.
	
	\item[(C$_4$)] (Smoothness) The spreading rates are sufficiently smooth. Technically speaking, $f_i^U$, $f_i^T$, $g_i^U$ and $g_i^R$ are twice continuously differentible.
	
	\item[(C$_5$)] (Monotonicity) The spreading rates are strictly increasing with respect to every relevant argument. That is,  $\frac{\partial f_i^U(x_1, \cdots, x_N)}{\partial x_j} > 0$ if $f_i^U$ is dependent upon $x_j$, $\frac{\partial f_i^T(x_1, \cdots, x_N)}{\partial x_j} > 0$ if $f_i^T$ is dependent upon $x_j$, $\frac{\partial g_i^U(x_1, \cdots, x_N)}{\partial x_j} > 0$ if $g_i^U$ is dependent upon $x_j$, and $\frac{\partial g_i^R(x_1, \cdots, x_N)}{\partial x_j} > 0$ if $g_i^R$ is dependent upon $x_j$.
	
	\item[(C$_6$)] (Concavity) The spreading rates flatten out and tend to saturation. That is, $\frac{\partial^2 f_i^U(x_1, \cdots, x_N)}{\partial x_j\partial x_k}\leq 0$, $\frac{\partial^2 f_i^T(x_1, \cdots, x_N)}{\partial x_j\partial x_k}\leq 0$, $\frac{\partial^2 g_i^U(x_1, \cdots, x_N)}{\partial x_j\partial x_k}\leq 0$, and  $\frac{\partial^2 g_i^R(x_1, \cdots, x_N)}{\partial x_j\partial x_k}\leq 0$.
	
\end{enumerate}

\noindent We refer to model (5) as the \emph{generic URTU model}. See Fig. 3 for the state transition rates of a person under the generic URTU model. Obviously, this model subsumes the linear URTU model as well as many other URTU models with nonlinear spreading rates.

\begin{figure}[!t]
	\centering
	\includegraphics[width=2.5in]{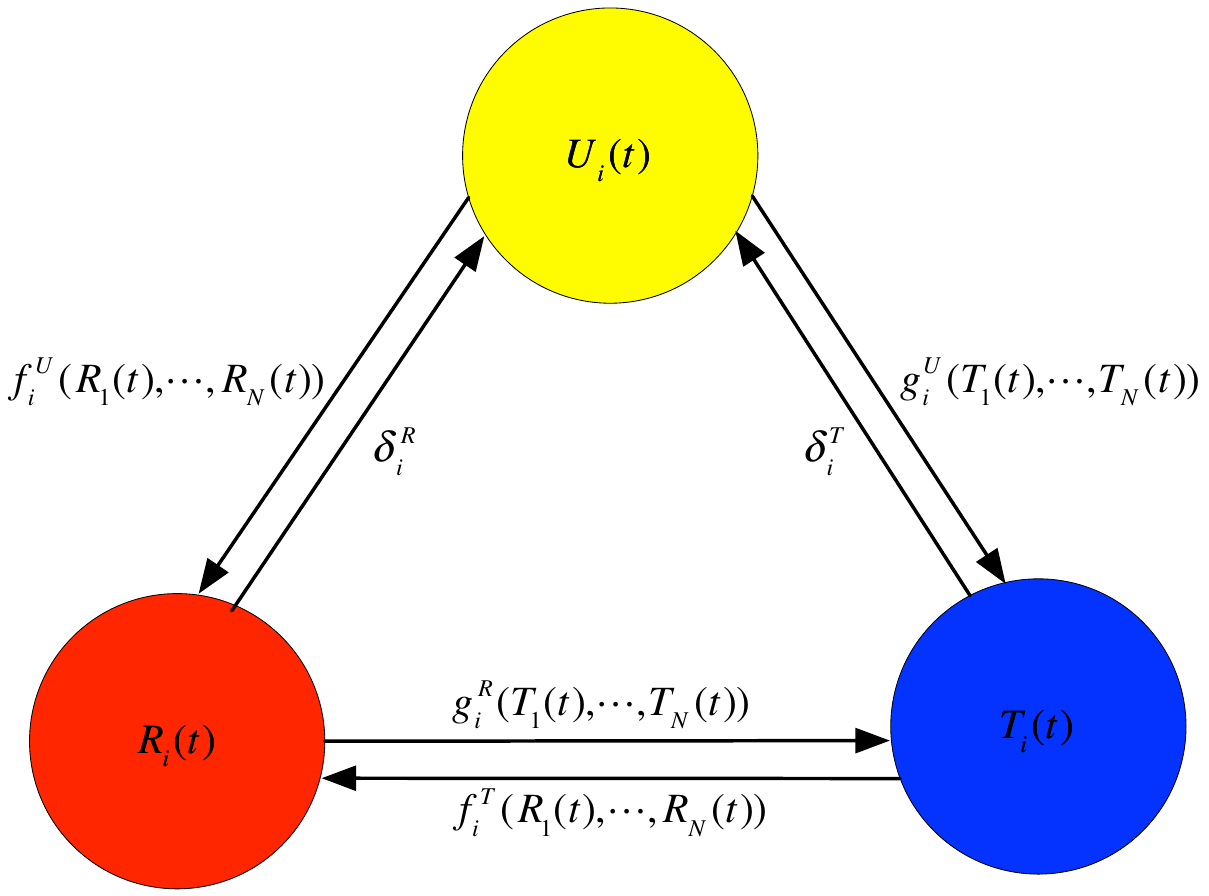}
	\caption{The state transition rates of person $i$ under the generic URTU model.}
\end{figure}

Let
\[
\Omega=\left\{(x_1,\cdots,x_{2N})\in \mathbb{R}_+^{2N}\mid x_i+x_{N+i}\leq 1, 1 \leq i \leq N\right\}.
\]
The initial state of model (5) lies in $\Omega$. It is easily shown that $\Omega$ is positively invariant for model (5).

Let us introduce the following matrix-vector notations.
\[
\mathbf{R}(t) = (R_1(t), \cdots, R_N(t))^T, \quad \mathbf{T}(t) = (T_1(t), \cdots, T_N(t))^T,
\]
\[
diag\mathbf{R}(t) = diag\left(R_i(t)\right), \quad diag\mathbf{T}(t) = diag\left(T_i(t)\right),
\]
\[
\begin{split}
\mathbf{f}_U(\mathbf{R}(t)) &= \left(f_1^U(\mathbf{R}(t)), \cdots, f_N^U(\mathbf{R}(t))\right)^T, \quad
\mathbf{f}_T(\mathbf{R}(t)) = \left(f_1^T(\mathbf{R}(t)), \cdots, f_N^T(\mathbf{R}(t))\right)^T,\\
\mathbf{g}_U(\mathbf{T}(t)) &= \left(g_1^U(\mathbf{T}(t)), \cdots, g_N^U(\mathbf{T}(t))\right)^T, \quad
\mathbf{g}_R(\mathbf{T}(t)) = \left(g_1^R(\mathbf{T}(t)), \cdots, g_N^R(\mathbf{T}(t))\right)^T.
\end{split}
\]
Then the generic URTU model can be written as
\begin{small}
	\begin{equation}
	\left\{
	\begin{aligned}
	\frac{d\mathbf{R}(t)}{dt}&=(\mathbf{E}_N-diag\mathbf{R}(t)-diag\mathbf{T}(t))\mathbf{f}_U(\mathbf{R}(t))-\mathbf{D}_R\mathbf{R}(t) +diag\mathbf{T}(t)\mathbf{f}_T(\mathbf{R}(t))- diag\mathbf{R}(t)\mathbf{g}_R(\mathbf{T}(t)),\\
	\frac{d\mathbf{T}(t)}{dt}&=(\mathbf{E}_N-diag\mathbf{R}(t)-diag\mathbf{T}(t))\mathbf{g}_U(\mathbf{T}(t))-\mathbf{D}_T\mathbf{T}(t) +diag\mathbf{R}(t)\mathbf{g}_R(\mathbf{T}(t))-diag\mathbf{T}(t)\mathbf{f}_R(\mathbf{R}(t)),
	\end{aligned}
	\right.
	\end{equation}
\end{small}
where $\mathbf{E}_N$ stands for the identity matrix of order $N$.

\section{Dynamics of the generic URTU model}

Consider the generic URTU model (5). Let $R(t)$ and $T(t)$ denote the fraction at time $t$ of rumor-believers and truth-believers, respectively. That is,
\begin{equation}
R(t) = \frac{1}{N}\sum_{i=1}^N R_i(t), \quad T(t) = \frac{1}{N}\sum_{i=1}^N T_i(t).
\end{equation}
The main aim of this work is to determine the developing tendency of $R(t)$ and $T(t)$ over time. For that purpose, we need some preliminary knowledges, which are listed below.

\subsection{Preliminaries}

For fundamental knowledge on matrix theory, see Ref. \cite{Horn2013}. In what follows, we consider only real square matrices. Given a matrix $\mathbf{A}$, let $s(\mathbf{A})$ denote the maximum real part of an eigenvalue of $\mathbf{A}$, and let $\rho(\mathbf{A})$ denote the spectral radius of $\mathbf{A}$, i.e., the maximum modulus of an eigenvalue of $\mathbf{A}$. $\mathbf{A}$ is \emph{Metzler} if its off-diagonal entries are all nonnegative.

%\begin{lm}(Reyleigh Formula, see a corollary of Theorem 4.2.2 in \cite{Horn2013})
%	Let $\mathbf{A}$ be a real symmetric matrix of order $n$, $\lambda_{\max}(\mathbf{A})$ the maximum eigenvalue of $\mathbf{A}$. Then,
%	\begin{equation*}
%	\lambda_{\max}(\mathbf{A}) = \max_{\mathbf{x} \in \mathbb{R}^n, \text{ } \mathbf{x} \neq \mathbf{0}} \frac{\mathbf{x}^T\mathbf{A}\mathbf{x}}{\mathbf{x}^T\mathbf{x}}.
%	\end{equation*}
%	Moreover, $\lambda_{\max}(\mathbf{A}) = \frac{\mathbf{x}^T\mathbf{A}\mathbf{x}}{\mathbf{x}^T\mathbf{x}}$ if and only if $\mathbf{x}$ is an eigenvector belonging to $\lambda_{\max}(\mathbf{A})$.
%\end{lm}

%\begin{lm}(Corollary 8.1.19 in \cite{Horn2013})
%	Let $\mathbf{A}$, $\mathbf{B}$ be a pair of nonnegative matrices of the same order. If $\mathbf{A} \leq \mathbf{B}$, then $\rho(\mathbf{A}) \leq \rho(\mathbf{B})$.
%\end{lm}

\begin{lm}(Corollary 8.1.30 in \cite{Horn2013})
	Let $\mathbf{A}$ be a nonnegative matrix. If $\mathbf{A}$ has a positive eigenvector $\mathbf{x}$, then $\rho(\mathbf{A})$ is an eigenvalue of $\mathbf{A}$, and $\mathbf{x}$ belongs to $\rho(\mathbf{A})$.
\end{lm}

%\begin{lm} \emph{(Theorem 8.4.4 in \cite{Horn2013})}
%	Let  $\mathbf{A}$ be an irreducible nonnegative matrix. Then (a) $\rho(\mathbf{A}) > 0$, (b) $\rho(\mathbf{A})$ is a simple eigenvalue of $\mathbf{A}$, and (c) up to scalar multiple, $\mathbf{A}$ has a unique positive eigenvector belonging to $\rho(\mathbf{A})$.
%\end{lm}

%This lemma is referred to as the \emph{Perron-Frobenius Theorem}.

\begin{lm}(Lemma 2.3 in \cite{Varga2000})
	Let $\mathbf{A}$ be an irreducible Metzler matrix. Then $s(\mathbf{A})$ is a simple eigenvalue of $\mathbf{A}$, and, up to scalar multiple, $\mathbf{A}$ has a unique positive eigenvector $\mathbf{x}$ belonging to $s(\mathbf{A})$.
\end{lm}

A matrix $\mathbf{A}$ is \emph{Hurwitz stable} or simply \emph{Hurwitz} if its eigenvalues all have negative real parts, i.e., $s(\mathbf{A}) < 0$.

\begin{lm}(Chapter 2 in \cite{Horn1991})
	A matrix $\mathbf{A}$ is Hurwitz if and only if there is a positive definite matrix $\mathbf{P}$ such that $ \mathbf{A}^T\mathbf{P}+\mathbf{P}\mathbf{A}$ is negative definite.
\end{lm}

A matrix $\mathbf{A}$ is \emph{diagonally stable} if there is a positive definite diagonal matrix $\mathbf{D}$ such that $\mathbf{A}^T\mathbf{D}+\mathbf{D}\mathbf{A}$ is negative definite. Obviously, a diagonally stable matrix is Hurwitz.

\begin{lm}(Section 2 in \cite{Narendra2010})
	A Metzler matrix is diagonally stable if it is Hurwitz.
\end{lm}

\begin{lm}(Lemma A.1 in \cite{Khanafer2014a})
	Let $\mathbf{A}$ be an irreducible Metzler matrix. If $s(\mathbf{A}) = 0$, then there is a positive definite diagonal matrix $\mathbf{D}$ such that $\mathbf{A}^T\mathbf{D}+\mathbf{D}\mathbf{A}$ is negative semi-definite.
\end{lm}

For fundamental theory on differential dynamical systems, see Ref. \cite{Khalil2002}.

\begin{lm} (Chaplygin Lemma, see Theorem 31.4 in \cite{Szarski1965}) Consider a smooth $n$-dimensional system of differential equations
	\[
	\frac{d\mathbf{x}(t)}{dt} = \mathbf{f}(\mathbf(\mathbf{x}(t)), \quad t \geq 0
	\]
	and the corresponding system of differential inequalities
	\[
	\frac{d\mathbf{y}(t)}{dt} \leq \mathbf{f}(\mathbf(\mathbf{y}(t)), \quad t \geq 0
	\]
	with $\mathbf{x}(0) = \mathbf{y}(0)$. Suppose that for any $a_1, \cdots, a_n \geq 0$, there hold
	\[
	 f_i(x_1+a_1, \cdots, x_{i-1}+a_{i-1}, x_i, x_{i+1} + a_{i+1}, \cdots, x_n + a_n) \geq f_i(x_1, \cdots, x_n), \quad i = 1, \cdots, n.
	\]
	Then $\mathbf{y}(t) \leq \mathbf{x}(t), t \geq 0$.
\end{lm}

\begin{lm} (Strauss-Yorke Theorem, see Corollary 3.3 in \cite{Strauss1967}) Consider a differential dynamical system
	\[
	\frac{d\mathbf{x}(t)}{dt} = \mathbf{f}(\mathbf(\mathbf{x}(t)) + \mathbf{g}(t, \mathbf{x}(t)), \quad t \geq 0,
	\]
	with $\mathbf{g}(t, \mathbf{x}(t)) \rightarrow \mathbf{0}$ when $t \rightarrow \infty$. Let
	\[
	\frac{d\mathbf{y}(t)}{dt} = \mathbf{f}(\mathbf(\mathbf{y}(t)), \quad t \geq 0
	\]
	denote the limit system of this system. If the origin is a global attractor for the limit system, and every solution to the original system is bounded on $[0, \infty)$, then the origin is also a global attractor for the original system.
\end{lm}

For fundamental knowledge on fixed point theory, see Ref. \cite{Agarwal2001}.

\begin{lm} (Brouwer Fixed Point Theorem, see Theorem 4.10 in \cite{Agarwal2001}) Let $C \subset \mathbb{R}^n$ be nonempty, bounded, closed, and convex. Let $f: C \rightarrow C$ be a continuous function. Then $f$ has a fixed point.
\end{lm}

\subsection{The equilibria}

The first step to understanding the dynamics of a differential dynamical system is to examine all of its equilibria. The generic URTU model might admit four different types of equilibria, which are defined as follows.

\begin{de}
	Let $\mathbf{E} = (\mathbf{R}^T, \mathbf{T}^T)^T$ be an equilibrium of the generic URTU model.
	
	\begin{enumerate}
		
		\item[(a)] $\mathbf{E}$ is \emph{uncertain} if  $\mathbf{R} = \mathbf{T}=\mathbf{0}$, which stands for the steady OSN state in which all persons are uncertain almost surely.
		
		\item[(b)] $\mathbf{E}$ is \emph{rumor-dominant} if $\mathbf{R} \neq \mathbf{0}$ and $\mathbf{T} = \mathbf{0}$, which stands for a steady OSN state in which some persons believe the rumor with positive probability and no person believes the truth almost surely.
		
		\item[(c)] $\mathbf{E}$ is \emph{truth-dominant} if $\mathbf{R} = \mathbf{0}$ and $\mathbf{T} \neq \mathbf{0}$, which stands for a steady OSN state in which some persons believe the truth with positive probability and no person believes the rumor almost surely.
		
		\item[(d)] $\mathbf{E}$ is \emph{coexistent} if $\mathbf{R} \neq \mathbf{0}$ and $\mathbf{T} \neq \mathbf{0}$, which stands for a steady OSN state in which some persons believe the rumor with positive probability and some persons believe the truth with positive probability.
		
	\end{enumerate}
	
\end{de}

Obviously, the generic URTU model always admits the uncertain equilibrium $\mathbf{E}_U=(0,\cdots,0)^T$. Due to the complexity of the model, we are unable to figure out its other equilibria. For our purpose, define a pair of Metzler matrices as follows.
\begin{equation}
\mathbf{Q}_1=\frac{\partial \mathbf{f}_U(\mathbf{0})}{\partial \mathbf{x}}-\mathbf{D}_R,
\quad \mathbf{Q}_2=\frac{\partial \mathbf{g}_U(\mathbf{0})}{\partial\mathbf{x}}-\mathbf{D}_T,
\end{equation}
%\[
%\begin{split}
%\mathbf{Q}_1^*&=\frac{\partial \mathbf{f}_T(\mathbf{0})}{\partial \mathbf{x}}-\mathbf{D}_R-diag\left(\mathbf{g}_R(\mathbf{1})\right),\\ \mathbf{Q}_2^*&=\frac{\partial{\mathbf{g}_R(\mathbf{0})}}{\partial\mathbf{x}}-\mathbf{D}_T-diag\left(\mathbf{f}_T(\mathbf{1})\right),
%\end{split}
%\]
where $\frac{\partial \mathbf{f}_U(\mathbf{0})}{\partial \mathbf{x}}$ and $\frac{\partial \mathbf{g}_U(\mathbf{0})}{\partial \mathbf{x}}$ stand for the Jacobian matrix of $\mathbf{f}_U$ and  $\mathbf{g}_U$ evaluated at the origin, respectively. As $G_R$ and $G_T$ are strongly connected, the four matrices are all irreducible.

%$\frac{\partial \mathbf{g}_U(\mathbf{0})}{\partial \mathbf{x}}$ and $\frac{\partial {\mathbf{g}_R(\mathbf{0})}}{\partial \mathbf{x}}$

We are ready to present the following fundamental result about the equilibria of the generic URTU model.

\begin{thm}
	Consider model (5). The following claims hold.
	\begin{enumerate}
		\item [(a)] If
		$s(\mathbf{Q}_1)>0$, then the model admits a unique rumor-dominant equilibrium. Let $\mathbf{E}_R=(\mathbf{R}^T, \mathbf{0}^T)^T$ denote the equilibrium, then $\mathbf{0} < \mathbf{R} < \mathbf{1}$.
		\item [(b)] If
		$s(\mathbf{Q}_2)>0$, then the model admits a unique truth-dominant equilibrium. Let $\mathbf{E}_T = (\mathbf{0}^T, \mathbf{T}^T)^T$ denote the equilibrium, then $\mathbf{0} < \mathbf{T} < \mathbf{1}$.
		%\item [(c)] If $s(\mathbf{Q}_1^*)>0$ and $s(\mathbf{Q}_2^*)>0$,
		%then the model admits a coexistent equilibrium. Let $\mathbf{E}_C = (\mathbf{R}^T, \mathbf{T}^T)^T$ denote such an equilibrium, then $\mathbf{0} < \mathbf{R} < \mathbf{1}$, $\mathbf{0} < \mathbf{T} < \mathbf{1}$.
	\end{enumerate}
\end{thm}

The proof of the theorem is left to Appendix B. This theorem manifests that the existence and locations of equilibria of the generic URTU model are dependent in a complex way upon the basic parameters as well as the network structures.

\subsection{Attractivity analysis}

Now, let us examine the attractivity of the equilibria of the generic URTU model. First, we have the following criterion for the attractivity of the uncertain equilibrium.

\begin{thm}
	Consider model (5). Suppose $s(\mathbf{Q}_1)\leq 0$ and $s(\mathbf{Q}_2)\leq 0$. Then the uncertain equilibrium $\mathbf{E}_U$ attracts $\Omega$. Hence, $R(t) \rightarrow 0$ and $T(t) \rightarrow 0$ as $t \rightarrow \infty$.
\end{thm}

The proof of the theorem is left to Appendix C. This theorem has the following useful corollary.

\begin{cor}
	The uncertain equilibrium $\mathbf{E}_U$ of model (5) attracts $\Omega$ if one of the following conditions is satisfied.
	
	\begin{enumerate}
		
		\item[(a)] $\rho(\mathbf{Q}_1\mathbf{D}_R^{-1}+\mathbf{E}_N) < 1$, $\rho(\mathbf{Q}_2\mathbf{D}_T^{-1}+\mathbf{E}_N) < 1$.
		
		\item[(b)] $\rho(\mathbf{B}_U\mathbf{D}_R^{-1}) < 1$, $\rho(\mathbf{C}_U\mathbf{D}_T^{-1}) < 1$.
		
		\item[(c)] $\sum_{i=1}^{N}\beta_{ij}^U < \delta_j^R$, $\sum_{i=1}^{N}\gamma_{ij}^U < \delta_j^T$, $j = 1, 2 \cdots, N$.
		
		\item[(d)] $\sum_{j=1}^{N}\frac{\beta_{ij}^U}{\delta_j^R} < 1$, $\sum_{j=1}^{N}\frac{\gamma_{ij}^U}{\delta_j^T} < 1$, $i = 1, 2, \cdots, N$.
		
	\end{enumerate}
\end{cor}

The proof of this corollary is left to Appendix D. The following theorem offers a criterion for the global attractivity of the rumor-dominant equilibrium.

\begin{thm}
	Consider model (5). Suppose $s(\mathbf{Q}_1)> 0$ and $s(\mathbf{Q}_2)\leq 0$. Then the rumor-dominant equilibrium $\mathbf{E}_R$ attracts $\{(\mathbf{R},\mathbf{T})\in\Omega:\mathbf{R}\neq \mathbf{0}\}$. Hence, if $R(0) \neq 0$, then $R(t) \rightarrow R^*$ and $T(t) \rightarrow 0$ as $t \rightarrow \infty$.
\end{thm}

The proof of the theorem is left to Appendix E. In parallel, we have the following criterion for the attractivity of the truth-dominant equilibrium.

\begin{thm}
	Consider model (5). Suppose $s(\mathbf{Q}_1)\leq 0$ and $s(\mathbf{Q}_2)> 0$. Then the truth-dominant equilibrium $\mathbf{E}_T$ attracts $\{(\mathbf{R},\mathbf{T})\in\Omega:\mathbf{T}\neq \mathbf{0}\}$. Hence, if $T(0) \neq 0$, then $R(t) \rightarrow 0$ and $T(t) \rightarrow T^*$ as $t \rightarrow \infty$.
\end{thm}

The argument for the theorem is analogous to that for Theorem 3 and hence is omitted.

Theorems 2 and 4 demonstrate that when (a) $s(\mathbf{Q}_1)\leq 0$ and $s(\mathbf{Q}_2)\leq 0$, or (b) $s(\mathbf{Q}_1)\leq 0$, $s(\mathbf{Q}_2) > 0$ and $T(0) \neq 0$, the rumor would terminate. In practice, the following measures are recommended to inhibit rumors.

\begin{enumerate}
  \item[(a)] Enhance the truth-spreading rates by presenting a convincing truth elaboration as early as possible.
  \item[(b)] Reduce the rumor-spreading rates by pointing out irrational aspects of the rumor.
  \item[(c)] Expand channels of spreading truths such as mass media and official announcement.
  \item[(d)] Lessen channels of spreading rumors by improving the quality of people.
\end{enumerate}

\section{Accuracy of the linear URTU model}

As was mentioned in Section 2, the exact URTU model accurately captures the average dynamics of the rumor-truth interaction, while the linear URTU model is an approximation of the exact URTU model. A question arises naturally: under what conditions does the linear URTU model satisfactorily capture this averaged dynamics? This section is devoted to giving an answer to the question via computer simulations.

For the comparison purpose, we need to numerically solve the exact URTU model, because its closed-form solution is far beyond our reach. Based on the standard Gillespie algorithm for numerically solving continuous-time Markov chain models \cite{Gillespie1977}, we develop a numerical algorithm for solving the exact URTU model. The basic idea of the algorithm is to take the average of $M = 10^{4}$ sample paths of the original URTU model as the solution to the exact URTU model. In the following experiments, a randomly chosen person is initialized as rumor-believing, a randomly chosen person is initialized as truth-believing, and all the remaining persons in the OSN are initialized as uncertain.

Scale-free networks are a large class of networks having widespread applications \cite{Albert2002}. Take a randomly generated scale-free network with 100 nodes as the rumor-spreading network as well as the truth-spreading network. By taking random combinations of the parameters, we get 4096 pairs of linear and exact URTU models, which are divided into four collections: 94 pairs for each of which $R(t)$ and $T(t)$ approach zero simultaneously, 1764 pairs for each of which $R(t)$ approaches a nonzero value but $T(t)$ approaches zero, 1770 pairs for each of which $R(t)$ approaches zero but $T(t)$ approaches a nonzero value, and 468 pairs for each of which both $R(t)$ and $T(t)$ approach nonzero values. By observation, we find that, for each of the four collections of pairs, the way that the dynamics of a linear URTU model deviates from that of the paired exact URTU model is qualitatively similar. Figs. 4-7 give the comparison results of two pairs for each collection, respectively.

\begin{figure}[!t]
	\centering
	\includegraphics[width=3in]{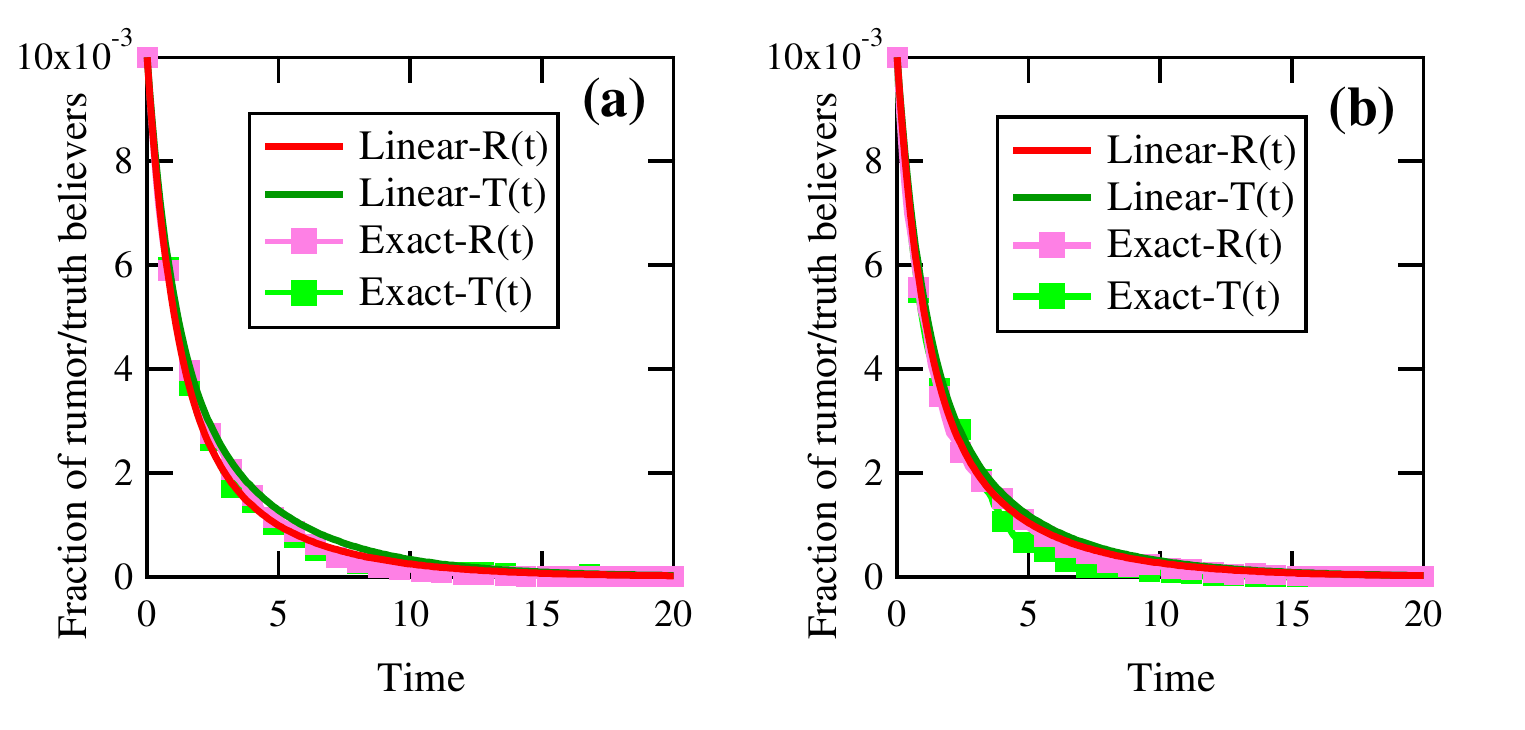}
	\caption{Comparison results for two pairs in the first collection.}
\end{figure}

\begin{figure}[!t]
	\centering
	\includegraphics[width=3in]{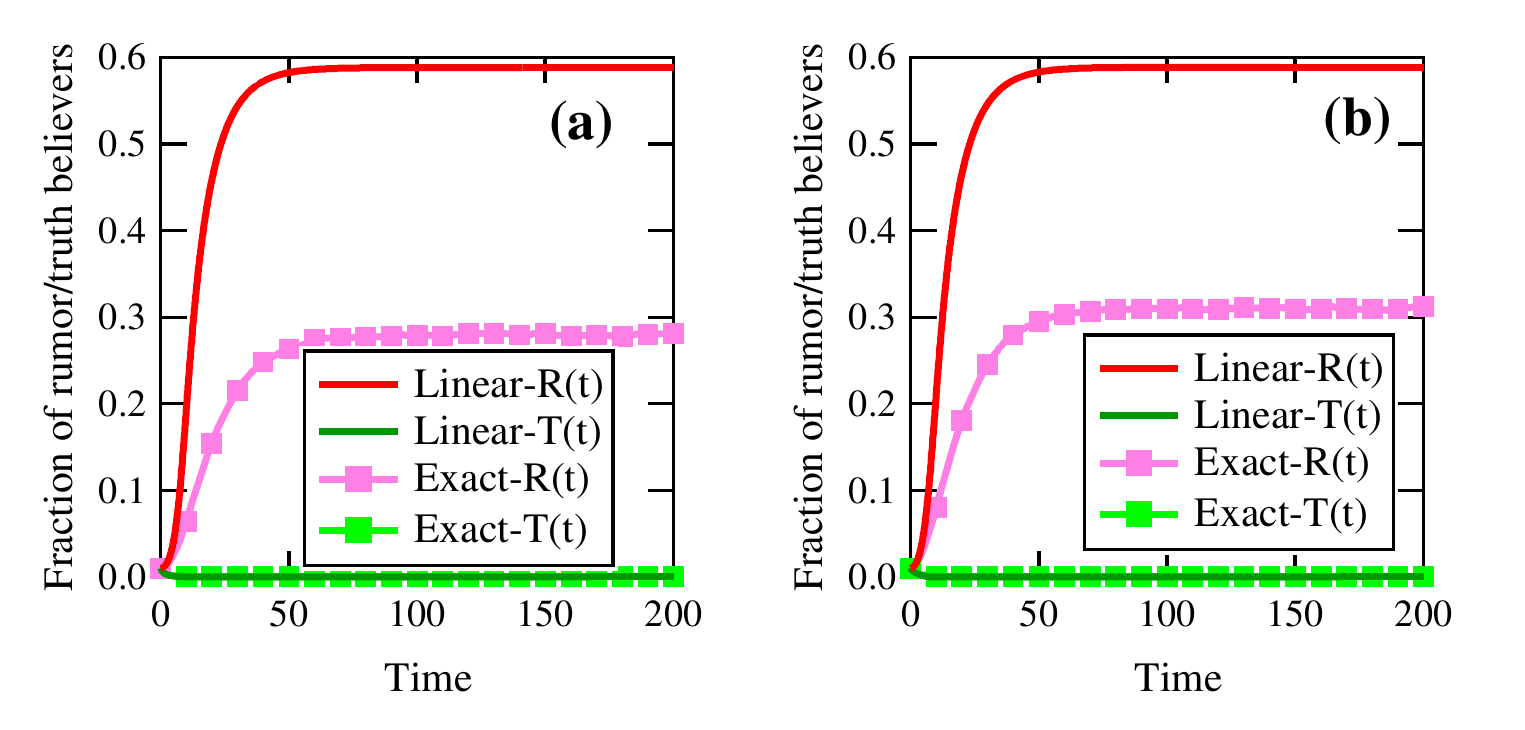}
	\caption{Comparison results for two pairs in the second collection.}
\end{figure}

\begin{figure}[!t]
	\centering
	\includegraphics[width=3in]{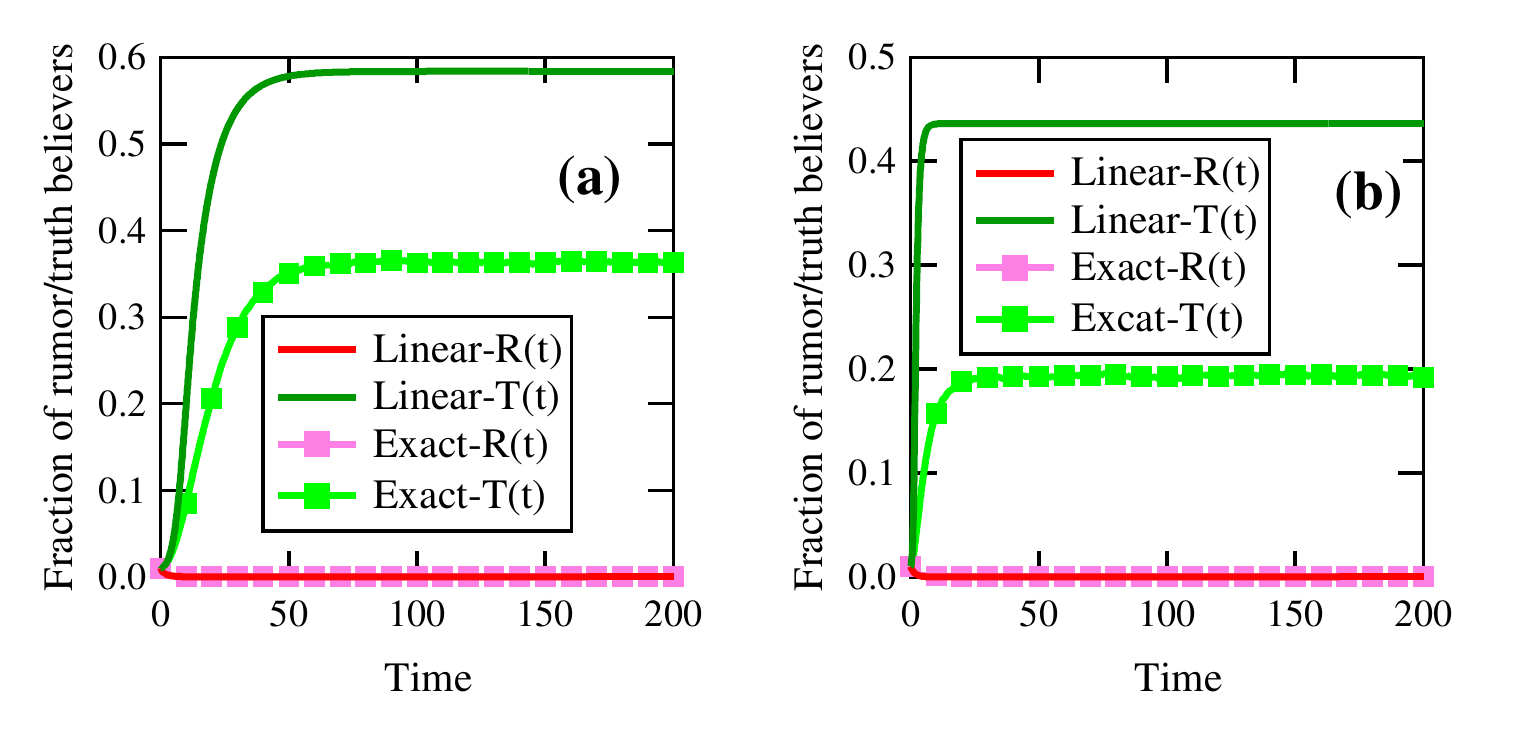}
	\caption{Comparison results for two pairs in the third collection.}
\end{figure}

\begin{figure}[!t]
	\centering
	\includegraphics[width=3in]{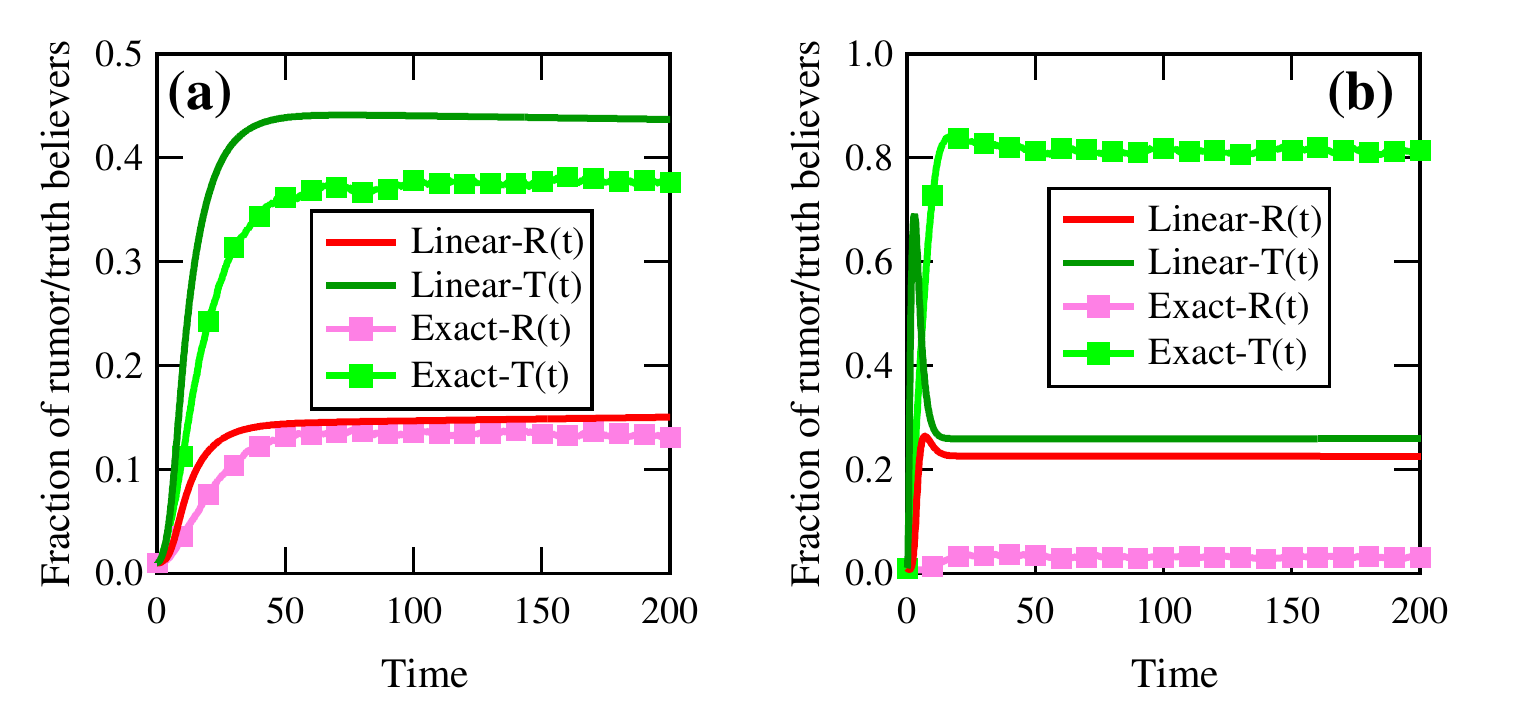}
	\caption{Comparison results for two pairs in the fourth collection.}
\end{figure}

Small-world networks are another large class of networks having widespread applications \cite{Watts1998}. Take a randomly generated small-word network with 100 nodes as the rumor-spreading network and the truth-distributing network. By taking random combinations of the parameters, we get 4096 pairs of linear and exact URTU models, which are divided into four collections: 151 pairs for each of which $R(t)$ and $T(t)$ approach zero simultaneously, 1657 pairs for each of which $R(t)$ approaches a nonzero value but $T(t)$ approaches zero, 1639 pairs for each of which $R(t)$ approaches zero but $T(t)$ approaches a nonzero value, and 649 pairs for each of which both $R(t)$ and $T(t)$ approach nonzero values. By observation, we find that, for each of the four collections of pairs, the way that the dynamics of a linear URTU model deviates from that of the paired exact URTU model is qualitatively similar. Figs. 8-11 give the comparison results of two pairs for each collection, respectively.

\begin{figure}[!t]
	\centering
	\includegraphics[width=3in]{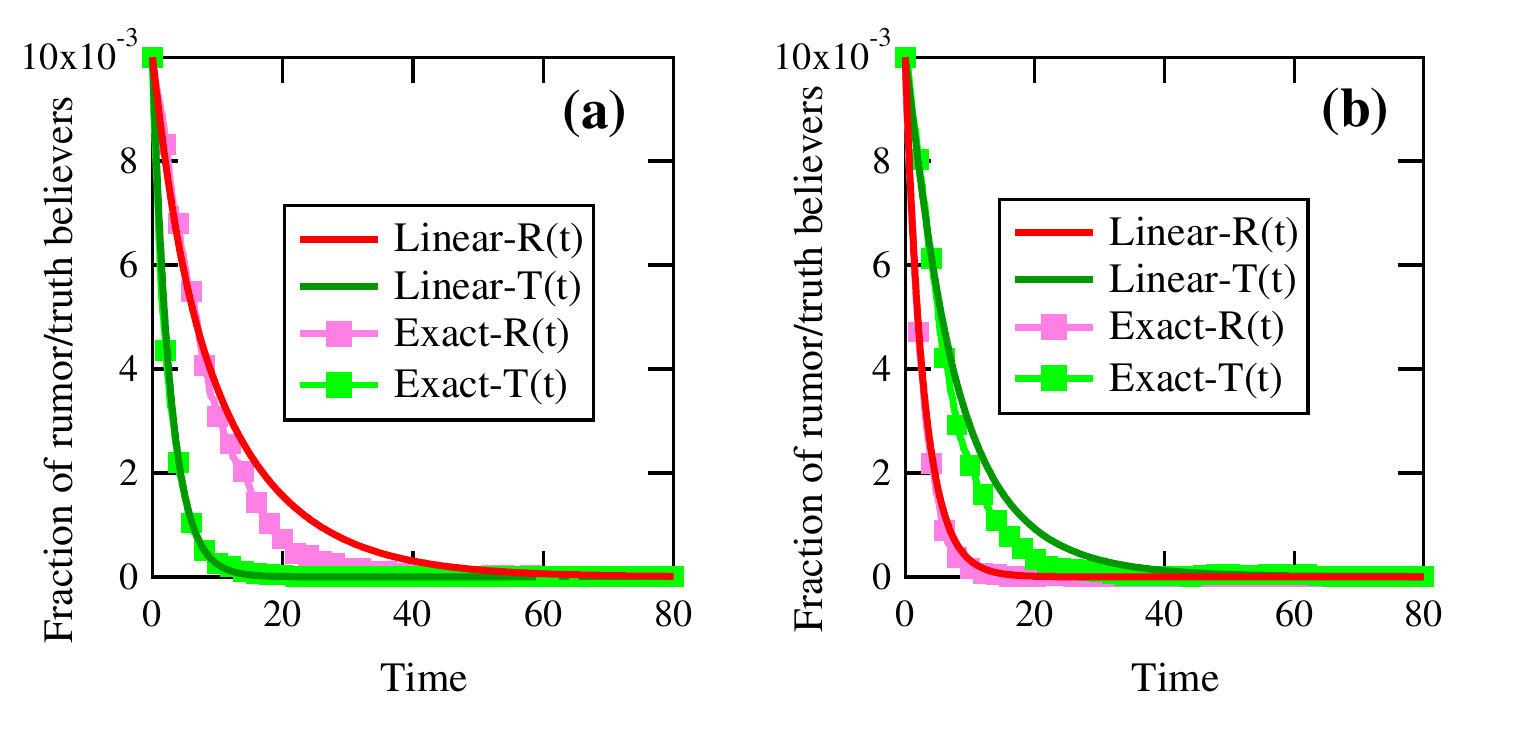}
	\caption{Comparison results for two pairs in the first collection.}
\end{figure}

\begin{figure}[!t]
	\centering
	\includegraphics[width=3in]{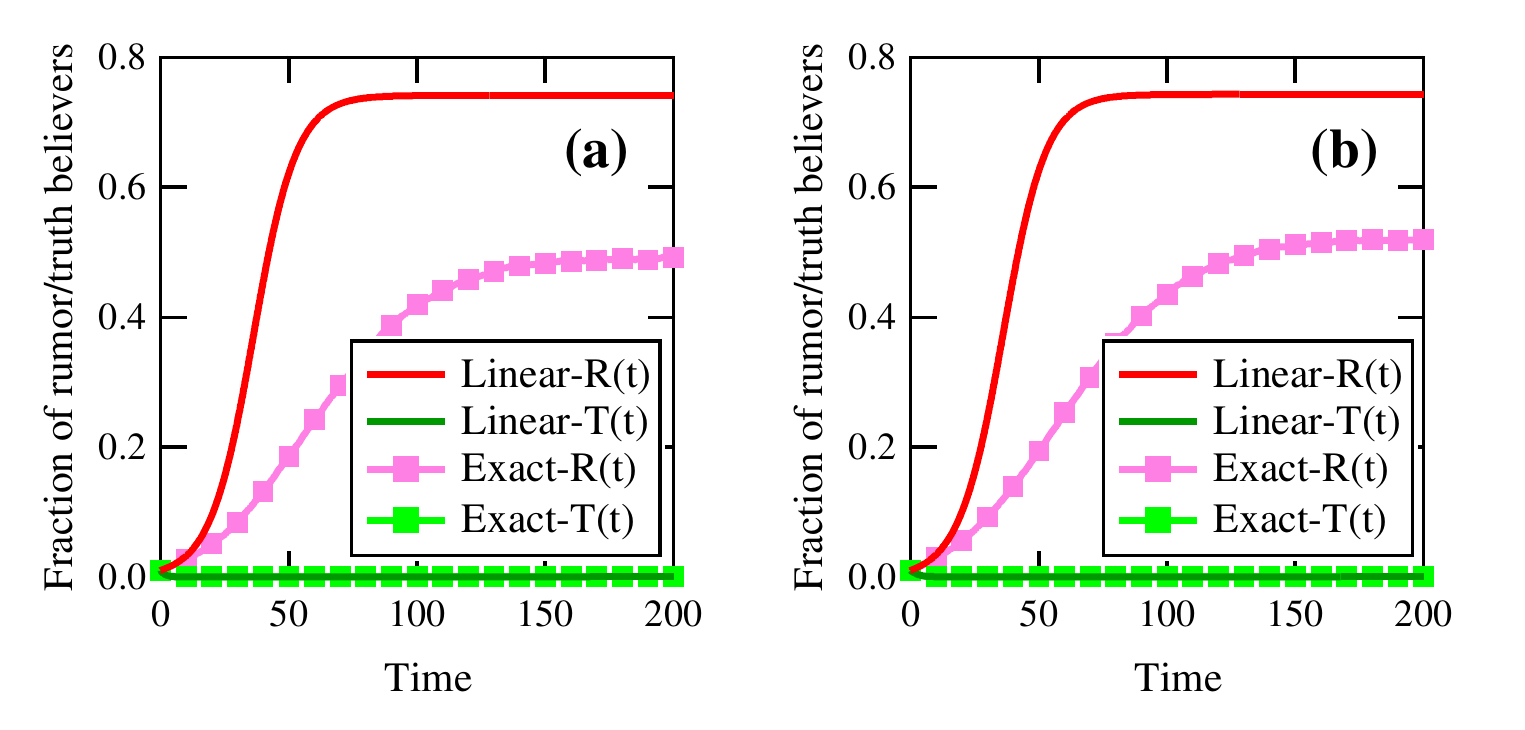}
	\caption{Comparison results for two pairs in the second collection.}
\end{figure}

\begin{figure}[!t]
	\centering
	\includegraphics[width=3in]{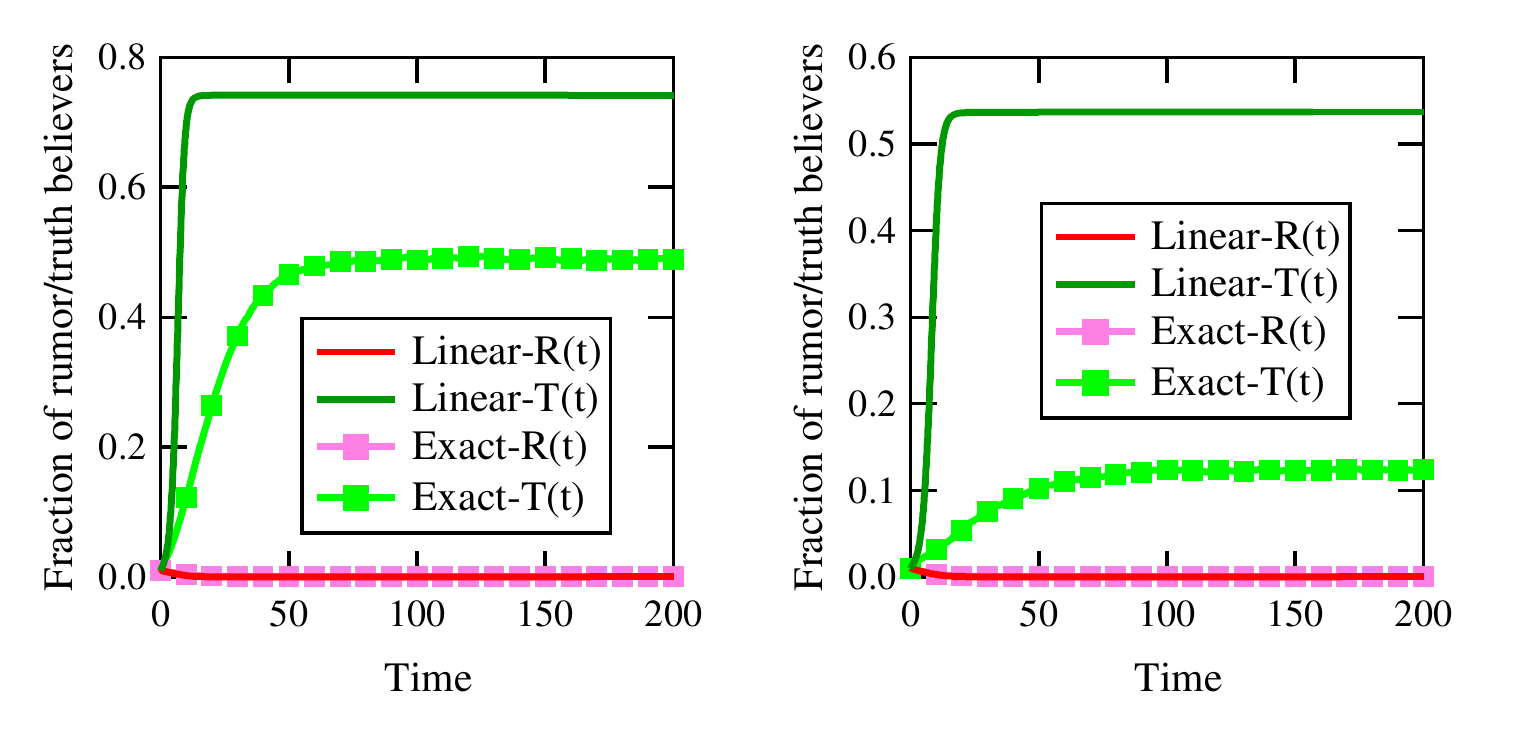}
	\caption{Comparison results for two pairs in the third collection.}
\end{figure}

\begin{figure}[!t]
	\centering
	\includegraphics[width=3in]{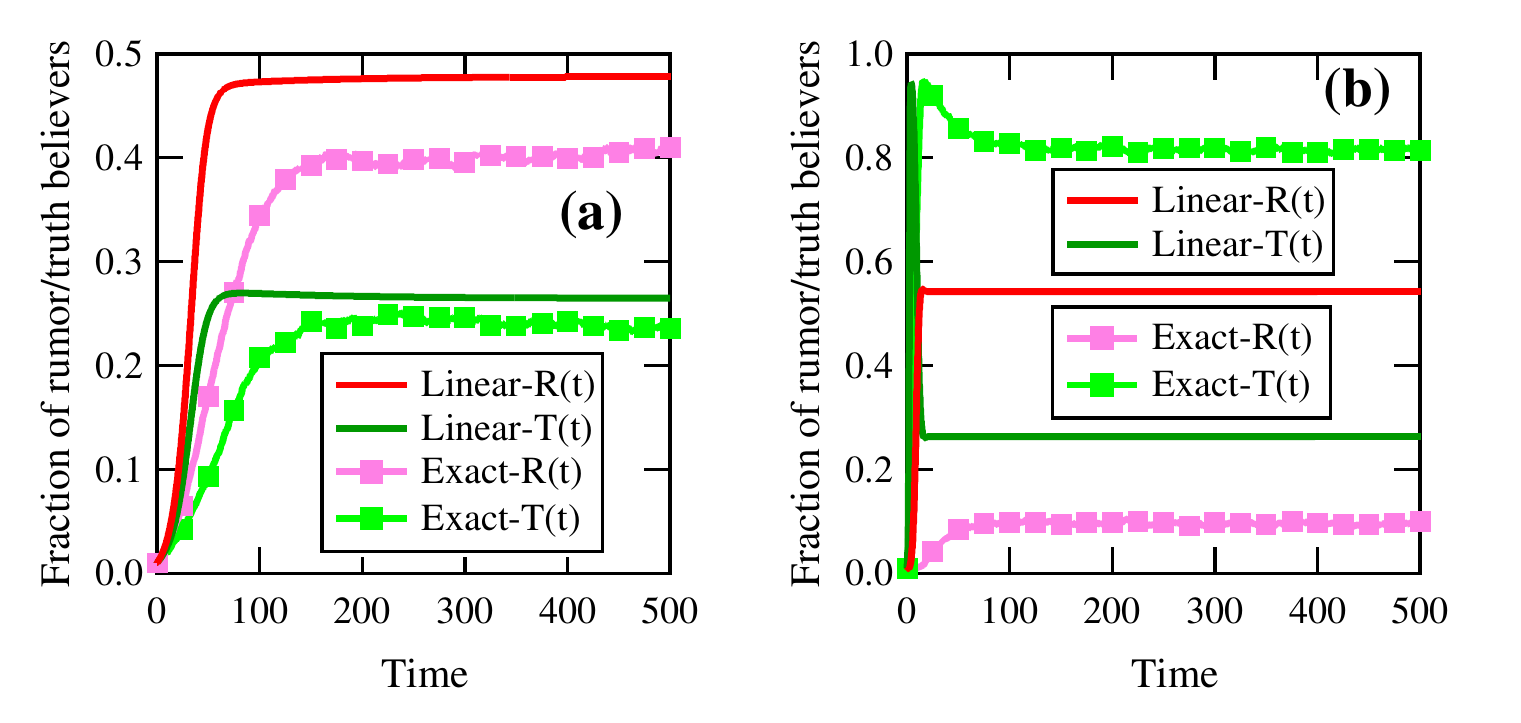}
	\caption{Comparison results for two pairs in the fourth collection.}
\end{figure}

The following conclusions are drawn from the previous examples.

\begin{enumerate}
	\item[(a)] If $R(t)$ approaches zero, then the linear URTU model accurately captures the average evolution process of the rumor. If $R(t)$ approaches a nonzero value, then the linear URTU model cannot accurately capture the average evolution process of the rumor.
	\item[(b)] If $T(t)$ approaches zero, then the linear URTU model accurately captures the average evolution process of the truth. If $T(t)$ approaches a nonzero value, then the linear URTU model cannot accurately capture the average evolution process of the truth.
\end{enumerate}

In the case where the linear URTU model works well, it can be employed to quickly predict the average evolution dynamics of the rumor or/and the truth in an OSN.

In the case where the linear URTU model doesn't work well, we have to resort to a generic URTU model with nonlinear spreading rates to achieve the goal of accurate prediction. In this case, the idea of deep learning might be employed to accurately estimate the spreading rates \cite{Goodfellow2016}.

\section{Concluding remarks}

This paper has discussed the effectiveness of the truth-spreading strategy for inhibiting rumors. A rumor-truth spreading model (the generic URTU model) is derived. Under the model, two criteria for the termination of a rumor have been presented. Extensive simulations show that, in some cases, the dynamics of a simplified URTU model (the linear URTU model) fits well with the actual rumor-truth interplay process. It is concluded that the generic URTU model (sometimes the linear URTU model) provides a theoretical basis for assessing the effectiveness of the truth-spreading strategy for restraining rumors.

Towards the direction, there are lots of works that are worth study. Under the generic URTU model, a criterion for the existence/attractivity of a coexistent equilibrium should be figured out, and the cost paid for restraining a rumor must be minimized \cite{YangLX2016, ZhangTR2017, BiJC2017}. Quarantining influential persons in an OSN who are spreading rumors is an effective measure of containing the prevalence of rumors other than spreading truths \cite{WenS2014}. As thus, it is valuable to develop a new rumor-truth spreading model that takes the quarantine effect into account. In the context of individual-level rumor-truth spreading models, it is of practical importance to understand the influence of more factors on the spread of rumors.

\section*{Acknowledgments}

This work is supported by Natural Science Foundation of China (Grant Nos. 61572006, 61379158),
National Sci-Tech Support Plan (Grant No. 2015BAF05B03), Natural Science Foundation of Chongqing (Grant
No. cstc2013jcyjA40011), and Fundamental Research Funds for the Central Universities (Grant No. 106112014CDJZR008823).


%\{References}

\appendix

\section{Proof of Lemma 1}

Given a sufficiently small time interval $\Delta t >0$, it follows from the total probability formula that
\[
\begin{split}
R_i(t+\Delta t)
&= \left(1 - R_i(t) - T_i(t)\right) \Pr\{X_{i}(t+\Delta t)=1 \mid X_{i}(t)=0\} + R_i(t) \Pr\{X_{i}(t+\Delta t)=1 \mid X_{i}(t)=1\} \\
& \quad + T_i(t) \Pr\{X_{i}(t+\Delta t)=1 \mid X_{i}(t)=2\}, \quad 1 \leq i \leq N. \quad\quad (*)
\end{split}
\]

By the conditional total probability formula and in view of model (1), we get that
\[
\begin{split}
\Pr\{X_{i}(t+\Delta t)=1 \mid X_{i}(t)=0\}
&=  \sum_{\mathbf{x} \in \{0, 1, 2\}^N, x_i = 0} \Pr\{X_{i}(t+\Delta t)=1 \mid X_{i}(t)=0, \mathbf{X}(t) = \mathbf{x}\} \cdot \Pr\{\mathbf{X}(t) = \mathbf{x} \mid X_{i}(t)=0\} \\
&= \frac{\Delta t}{1 - R_i(t) - T_i(t)} \cdot \sum_{\mathbf{x} \in \{0, 1, 2\}^N, x_i = 0} \sum_{j = 1}^N \beta_{ij}^U 1_{\{x_j=1\}} \cdot \Pr\{\mathbf{X}(t) = \mathbf{x}\} + o(\Delta t) \\
&= \frac{\Delta t}{1 - R_i(t) - T_i(t)} \cdot \sum_{j = 1}^N \beta_{ij}^U \sum_{\mathbf{x} \in \{0, 1, 2\}^N} 1_{\{x_i = 0, x_j=1\}} \cdot \Pr\{\mathbf{X}(t) = \mathbf{x}\} + o(\Delta t) \\
&= \frac{\Delta t}{1 - R_i(t) - T_i(t)}  \sum_{j = 1}^N \beta_{ij}^U \Pr\{X_i(t) = 0, X_j(t) = 1\} + o(\Delta t), \quad 1 \leq i \leq N.
\end{split}
\]

\noindent Similarly, we can derive that
\[
\begin{split}
\Pr\{X_{i}(t+\Delta t)=2 \mid X_{i}(t)=1\}
&= \frac{\Delta t}{R_i(t)} \cdot \sum_{j = 1}^N \gamma_{ij}^R \Pr\{X_i(t) = 1, X_j(t) = 2\} + o(\Delta t), \quad 1 \leq i \leq N, \\
\Pr\{X_{i}(t+\Delta t) = 0 \mid X_{i}(t)=1\} &= \delta_i^R \Delta t+o(\Delta t), \quad 1 \leq i \leq N.
\end{split}
\]

\noindent It follows that
\[
\Pr\{X_{i}(t+\Delta t)=1 \mid X_{i}(t)=1\}
= 1-\frac{\Delta t}{R_i(t)} \cdot \sum_{j = 1}^N \gamma_{ij}^R \Pr\{X_i(t) = 1, X_j(t) = 2\} -\delta_i^ R \Delta t + o(\Delta t), \quad 1 \leq i \leq N.
\]

\noindent Besides, we have
\[
\Pr\{X_{i}(t+\Delta t)=1 \mid X_{i}(t)=2\}
= \frac{\Delta t}{T_i(t)} \cdot \sum_{j = 1}^N \beta_{ij}^T \Pr\{X_i(t) = 2, X_j(t) = 1\} + o(\Delta t), \quad 1 \leq i \leq N.
\]

\noindent Substituting these equations into Eqs. (*), rearranging the terms, dividing both sides by $\Delta t$, and letting $\Delta t \rightarrow 0$, we get that, for $1 \leq i \leq N$,
\[
\begin{split}
\frac{dR_i(t)}{dt} &= \sum_{j = 1}^N \beta_{ij}^U \Pr\{X_i(t) = 0, X_j(t) = 1\} +\sum_{j = 1}^N \beta_{ij}^T \Pr\{X_i(t) = 2, X_j(t) = 1\} \\
& \quad - \sum_{j = 1}^N \gamma_{ij}^R \Pr\{X_i(t) = 1, X_j(t) = 2\} - \delta_i^R R_i(t), \quad 1 \leq i \leq N.
\end{split}
\]

\noindent Similarly, we can derive the last $N$ equations in Lemma 1. The proof is complete.

\section{Proof of Theorem 1}
(a) Suppose model (5) admits a rumor-dominant equilibrium
$\mathbf{E}=(R_1,\cdots,R_N,0,\cdots, 0)^T$. Let $\mathbf{R}=(R_1,\cdots,R_N)^T$. We show that $\mathbf{0} < \mathbf{R} < \mathbf{1}$. It follows from the model that
\[
R_i=\frac{f_i^U(\mathbf{R})}{\delta_i^R+f_i^U(\mathbf{R})}< 1, \quad 1 \leq i \leq N.
\]
Hence, $\mathbf{R} < \mathbf{1}$. On the contrary, suppose that some $R_k = 0$. It follows from model (5) that $f_k^U(\mathbf{R})=0$. As $G_R$ is strongly connected, we get that some $\beta_{kl}^U >0$, implying that $R_l = 0$. Repeating this argument, we finally get that $\mathbf{R} = \mathbf{0}$, contradicting the assumption that  $\mathbf{E}$ is a rumor-dominant equilibrium. Hence, $\mathbf{R} > \mathbf{0}$.

Define a continuous mapping $\mathbf{H}=(H_1,\cdots,H_N)^T: (0,1]^N \rightarrow (0,1]^N$ by
\[
H_i(\mathbf{x})=\frac{f_i^U(\mathbf{x})}{\delta_i^R+f_i^U(\mathbf{x})}, \quad \mathbf{x}=(x_1,\cdots,x_N)^T\in(0,1]^N.
\]
It suffices to show that $\mathbf{H}$ has a unique fixed point.
Let $\mathbf{T}(t) \equiv \mathbf{0}$ and rewrite model (5) as
\[
\frac{d\mathbf{R}(t)}{dt}=\mathbf{Q}_1\mathbf{R}(t)+\mathbf{G}(\mathbf{R}(t)),
\]
where $\mathbf{G}(\mathbf{R}(t))=o(\|\mathbf{R}(t)\|)$. By Lemma 3, $\mathbf{Q}_1$ has a positive eigenvector $\mathbf{v}=(v_1,\cdots,v_N)^T$ belonging to the eigenvalue $s(\mathbf{Q}_1)$. As $s(Q_1) > 0$, we have
$\mathbf{Q}_1\mathbf{v}=s(\mathbf{Q}_1)\mathbf{v} > \mathbf{0}.$
Hence, there is a small $\varepsilon > 0$ such that
\[\mathbf{Q}_1 \cdot (\varepsilon \mathbf{v})+\mathbf{G}(\varepsilon \mathbf{v})=\varepsilon s(\mathbf{Q}_1)\mathbf{v}+\mathbf{G}(\varepsilon \mathbf{v})\geq \mathbf{0},
\]
which is equivalent to $\mathbf{H}(\varepsilon \mathbf{v})\geq \varepsilon \mathbf{v}$. On the other hand, it is easily verified that $\mathbf{H}$ is monotonically increasing, i.e., $\mathbf{u} \geq \mathbf{w}$ implies $\mathbf{H}(\mathbf{u}) \geq \mathbf{H}(\mathbf{w})$.  Define a compact convex set as
\[
K=\prod_{i=1}^N[\varepsilon v_i,1].
\]
Then $\mathbf{H}|_{K}$
maps $K$ into $K$. It follows from Lemma 9 that $\mathbf{H}$ has a fixed point in $K$. Denote this fixed point by $\mathbf{R}^*=(R_1^*,\cdots,R_N^*)^T$

Suppose $\mathbf{H}$ has a fixed point $\mathbf{R}^{**}=(R_1^{**},\cdots,R_N^{**})^T$ other than $ \mathbf{R}^{*}$. Let
\[
\theta=\max_{1 \leq i \leq N}\frac{R_i^*}{R_i^{**}}, \quad i_0=\arg\max_{1 \leq i \leq N}\frac{R_i^*}{R_i^{**}}.
\]
Without loss of generality, assume $\theta>1$. It follows that
\[
R_{i_0}^{*}= H_{i_0}(\mathbf{R}^*)\leq H_{i_0}(\theta\mathbf{R}^{**}) \frac{f_{i_0}^U(\theta\mathbf{R}^{**})}{\delta_{i_0}^R+f_{i_0}^U(\theta\mathbf{R}^{**})}< \frac{f_{i_0}^U(\theta \mathbf{R}^{**})}{\delta_{i_0}^R+f_{i_0}^U(\mathbf{R}^{**})}
\leq \frac{\theta f_{i_0}^U(\mathbf{R}^{**})}{\delta_{i_0}^R+f_{i_0}^U(\mathbf{R}^{**})}=\theta H_{i_0}(\mathbf{R}^{**})=\theta R_{i_0}^{**},
\]
where $f_{i_0}^U(\theta\mathbf{R}^{**}) \leq \theta f_{i_0}^U(\mathbf{R}^{**})$ follows from the concavity of $f_{i_0}^U$. This contradicts the assumption that $R_{i_0}^{*}=\theta R_{i_0}^{**}$. Hence, $\mathbf{R}^{*}$ is the unique fixed point of $\mathbf{H}$. The proof is complete.

(b) The argument is analogous to that for Claim (a) and hence is
omitted.

\section{Proof of Theorem 2}
Let $(\mathbf{R}(t)^T, \mathbf{T}(t)^T)^T$ be a solution to model (5). It follows from the first $N$ equations of model (6) that
\[\frac{d\mathbf{R}(t)}{dt}\leq (\mathbf{E}_N-diag\mathbf{R}(t)) \mathbf{f}_U (\mathbf{R}(t))-\mathbf{D}_R \mathbf{R}(t).
\]
Consider the comparison system
\[
\frac{d\mathbf{u}(t)}{dt}=(\mathbf{E}_N-diag\mathbf{u}(t))\mathbf{f}_U(\mathbf{u}(t))-\mathbf{D}_R \mathbf{u}(t), \quad \quad (**)
\]
with $\mathbf{u}(0) = \mathbf{R}(0)$. This system admits the trivial equilibrium $\mathbf{0}$. Moreover, it follows from Lemma 7 that $\mathbf{u}(t) \geq \mathbf{R}(t) \geq \mathbf{0}$. We proceed by distinguishing two possibilities.

\emph{Case 1}: $s(\mathbf{Q}_1)<0$. By Lemma 5, there is a positive definite diagonal matrix $\mathbf{P}_1$ such that $\mathbf{Q}_1^T\mathbf{P}_1+\mathbf{P}_1\mathbf{Q}_1$ is negative definite. Let $\mathbf{u} = (u_1,\cdots,u_N)^T$, and define a
positive definite function as
\[
V_1(\mathbf{u})=\mathbf{u}^T \mathbf{P}_1 \mathbf{u}.
\]
By calculations, we get
\[
\begin{split}
\frac{dV_1(\mathbf{u}(t))}{dt}\mid_{(**)} &= 2\mathbf{u}(t)^T \mathbf{P}_1 \frac{d\mathbf{u}(t)}{dt}\leq 2\mathbf{u}(t)^T \mathbf{P}_1\left[\mathbf{f}_U(\mathbf{u}(t)) - \mathbf{D}_R\mathbf{u}(t) \right] \leq 2\mathbf{u}(t)^T \mathbf{P}_1\mathbf{Q}_1 \mathbf{u}(t)\\
&= \mathbf{u}(t)^T[\mathbf{Q}_1^T\mathbf{P}_1+\mathbf{P}_1\mathbf{Q}_1]\mathbf{u}(t) \leq 0.
\end{split}
\]

\noindent Here the second inequality follows from the concavity of $\mathbf{f}_U(\mathbf{x}) - \mathbf{D}_R\mathbf{x}$. Furthermore,
$\frac{dV_1(\mathbf{u}(t))}{dt}\mid_{(**)}=0$ if and
only if $\mathbf{u}(t)=\mathbf{0}$. According to the
LaSalle Invariance Principle (Corollary 4.1 in \cite{Khalil2002}), the trivial equilibrium $\mathbf{0}$ of system (**) is asymptotically stable for $[0,1]^N$.

\emph{Case 2}: $s(\mathbf{Q}_1)=0$. By Lemma 6, there is a positive definite diagonal matrix $\mathbf{P}_2$ such that $\mathbf{Q}_1^T\mathbf{P}_2+\mathbf{P}_2\mathbf{Q}_1$ is negative semi-definite. Define a
positive definite function as
\[
V_2(\mathbf{u})=\mathbf{u}^T \mathbf{P}_2 \mathbf{u}.
\]
Similarly, we have
\[
\frac{dV_2(\mathbf{u}(t))}{dt}\mid_{(**)}
\leq \mathbf{u}(t)^T[\mathbf{Q}_1^T\mathbf{P}_2+\mathbf{P}_2\mathbf{Q}_1]\mathbf{u}(t) \leq 0.
\]
If $\mathbf{Q}_1^T\mathbf{P}_2+\mathbf{P}_2\mathbf{Q}_1$ is negative definite, the subsequent argument is analogous to that for Case 1. Now, assume $\mathbf{Q}_1^T\mathbf{P}_2+\mathbf{P}_2\mathbf{Q}_1$ is not negative definite, which implies
\[
  s(\mathbf{Q}_1^T\mathbf{P}_2+\mathbf{P}_2\mathbf{Q}_1)=0.
\]
As $\mathbf{Q}_1^T\mathbf{P}_2+\mathbf{P}_2\mathbf{Q}_1$ is Metzler and irreducible, it follows from Lemma 3 that (a) $0$ is a simple eigenvalue of $\mathbf{Q}_1^T\mathbf{P}_2+\mathbf{P}_2\mathbf{Q}_1$, and (b) up to scalar multiple, $\mathbf{Q}_1^T\mathbf{P}_2+\mathbf{P}_2\mathbf{Q}_1$ has a positive eigenvector belonging to eigenvalue 0. Obviously,
$\frac{dV_2(\mathbf{u}(t))}{dt}\mid_{(**)}=0$ if $\mathbf{u}(t)=\mathbf{0}$. On the contrary, suppose $\frac{dV_2(\mathbf{u}(t))}{dt}\mid_{(**)}=0$ for some $\mathbf{u}(t)\geq \mathbf{0}$. If $\mathbf{u}(t) > \mathbf{0}$, then $\mathbf{f}_U(\mathbf{u}(t))>\mathbf{0}$, implying $\frac{dV_2(\mathbf{u}(t))}{dt}\mid_{(**)}<0$, a contradiction. If $\mathbf{u}(t)$ has a zero component, then $\mathbf{u}(t)$ is not an eigenvector of $\mathbf{Q}_1^T\mathbf{P}_2+\mathbf{P}_2\mathbf{Q}_1$ belonging to eigenvalue 0. It follows from the Rayleigh formula (Theorem 4.2.2 in \cite{Horn2013}) that
\[
\mathbf{u}(t)^T[\mathbf{Q}_1^T\mathbf{P}_2+\mathbf{P}_2\mathbf{Q}_1]\mathbf{u}(t)<0,
\]
implying $\frac{dV_2(\mathbf{u}(t))}{dt}\mid_{(**)}<0$, again a contradiction. Hence, $\mathbf{u}(t)=\mathbf{0}$ if $\frac{dV_2(\mathbf{u}(t))}{dt}\mid_{(**)}=0$. It follows from the
LaSalle Invariance Principle that the trivial equilibrium $\mathbf{0}$ of system (**) is asymptotically stable with respect to $[0,1]^N$.

Combining Cases 1 and 2, we get $\mathbf{u}(t) \rightarrow \mathbf{0}$ as $t \rightarrow \infty$. According to Lemma 7, we get $\mathbf{R}(t)\leq \mathbf{u}(t)$, which implies $\mathbf{R}(t) \rightarrow \mathbf{0}$ as $t \rightarrow \infty$.

Similarly, we can derive that $\mathbf{T}(t) \rightarrow \mathbf{0}$ as $t \rightarrow \infty$. The proof is complete.

\section{Proof of Corollary 1}
(a) We first show $s(\mathbf{Q}_1) < 0$. As $\mathbf{Q}_1\mathbf{D}_R^{-1}$ is Metzler and irreducible, it follows from Lemma 3 that $\mathbf{Q}_1\mathbf{D}_R^{-1}$ has a positive eigenvector $\mathbf{x}$ belonging to eigenvalue $s(\mathbf{Q}_1\mathbf{D}_R^{-1})$. So,
\[
  (\mathbf{Q}_1\mathbf{D}_R^{-1}+\mathbf{E}_N)\mathbf{x}=[s(\mathbf{Q}_1\mathbf{D}_R^{-1})+1]\mathbf{x}.
\]
That is, $\mathbf{x}$ is an eigenvector of $\mathbf{Q}_1\mathbf{D}_R^{-1}+\mathbf{E}_N$ belonging to eigenvalue $s(\mathbf{Q}_1\mathbf{D}_R^{-1})+1$. It follows from Lemma 2 that
\[
  s(\mathbf{Q}_1\mathbf{D}_R^{-1}) = \rho(\mathbf{Q}_1\mathbf{D}_R^{-1}+\mathbf{E}_N)-1 < 0.
\]
By Lemma 5, there is a positive definite diagonal matrix $\mathbf{D}$ such that the matrix
\[
\mathbf{P}=(\mathbf{Q}_1\mathbf{D}_R^{-1})^T\mathbf{D}+\mathbf{D}(\mathbf{Q}_1\mathbf{D}_R^{-1})
\]
is negative definite. Direct calculations give
\[
\left[\mathbf{D}_R^{\frac{1}{2}}\mathbf{Q}_1\mathbf{D}_R^{-\frac{1}{2}}\right]^T\mathbf{D}+\mathbf{D}\left[\mathbf{D}_R^{\frac{1}{2}}\mathbf{Q}_1\mathbf{D}_R^{-\frac{1}{2}}\right]
=\mathbf{D}_R^{\frac{1}{2}}\mathbf{P}\mathbf{D}_R^{\frac{1}{2}}.
\]
As $\mathbf{D}_R^{\frac{1}{2}}\mathbf{P}\mathbf{D}_R^{\frac{1}{2}}$ is negative definite, $\mathbf{D}_R^{\frac{1}{2}}\mathbf{Q}_1\mathbf{D}_R^{-\frac{1}{2}}$ is diagonally stable and hence Hurwitz. It follows that
\[
  s(\mathbf{Q}_1) = s(\mathbf{D}_R^{\frac{1}{2}}\mathbf{Q}_1\mathbf{D}_R^{-\frac{1}{2}}) < 0.
\]

Similarly, we have $s(\mathbf{Q}_2) < 0$. The declared result follows from Theorem 2.

(b) By the concavity of $f_i^U(\mathbf{x})$, we have
$\frac{\partial f_i^U(\mathbf{0})}{\partial x_j}\leq \beta_{ij}^U$. That is, $\mathbf{Q}_1 + \mathbf{D}_R \leq \mathbf{B}_U$. Hence,
\[
  \rho(\mathbf{Q}_1\mathbf{D}_R^{-1}+\mathbf{E}_N) \leq \rho(\mathbf{B}_U\mathbf{D}_R^{-1}) < 1.
\]
Similarly, we have $\rho(\mathbf{Q}_2\mathbf{D}_T^{-1}+\mathbf{E}_N) < 1$. The claim follows from Claim (a) of this corollary.

(c) The claim follows from Claim (b) of this corollary and $\rho(\mathbf{M}) \leq ||\mathbf{M}||_1$.

(d) The claim follows from Claim (b) of this corollary and $\rho(\mathbf{M}) \leq ||\mathbf{M}||_{\infty}$.

\section{Proof of Theorem 3}
Let $(\mathbf{R}(t)^T, \mathbf{T}(t)^T)^T$ be a solution
to model (5) with $\mathbf{R}(0)\neq \mathbf{0}$. It follows from the last $N$ equations of model (5) that
\[
  \frac{d\mathbf{T}(t)}{dt}\leq(\mathbf{E}_N-diag(\mathbf{T}(t)))\mathbf{g}^U(\mathbf{T}(t))-\mathbf{D}_T\mathbf{T}(t).
\]
By an argument analogous to that for Theorem 2, we get $\mathbf{T}(t)\rightarrow\mathbf{0}$. Consider the following limit system of model (5).
\[
  \frac{d\mathbf{u}(t)}{dt}=(\mathbf{E}_N-diag\mathbf{u}(t))\mathbf{f}(\mathbf{u}(t))-\mathbf{D}_R \mathbf{u}(t), \quad \quad (***)
\]
with $\mathbf{u}(0) = \mathbf{R}(0)$. Theorem 1 confirms that the system admits a unique nonzero equilibrium $\mathbf{R}^*=(R_1^*,\cdots,R_N^*)$. By Lemma 8, it suffices to show that $\mathbf{R}^*$ is asymptotically stable for $(0,1]^N$. Given a solution $\mathbf{u}(t) = (u_1(t), \cdots, u_N(t))^T$ to system (***) with $\mathbf{u}(0) > \mathbf{0}$. First, let us show the following claim.

\emph{Claim 1:} $\mathbf{u}(t) > \mathbf{0}$ for all $t > 0$.

\emph{Proof of Claim 1:} On the contrary, suppose there is $t_0 > 0$ such that (a) $\mathbf{u}(t) > \mathbf{0}$, $0 < t < t_0$, and (b) $u_i(t_0)=0$ for some $i$. According to the smoothness of $\mathbf{u}(t)$, we get $\frac{du_{i}(t_0)}{dt}=0$, implying $f_{i}^U(\mathbf{u}(t_0))=0$. As $G_R$ is strongly connected, there is $j$ such that $\beta_{ij}^U>0$, which implies $u_{j}(t_0)=0$. Working inductively, we conclude that $\mathbf{u}(t_0)=0$. This contradicts the uniqueness of the solution to system (***) with given initial condition. Claim 1 is proven.

For $t > 0$, let
\[
Z(\mathbf{u}(t))=\max_{1 \leq k \leq N} \frac{u_k(t)}{R_k^*}, \quad
z(\mathbf{u}(t))=\min_{1 \leq k \leq N} \frac{u_k(t)}{R_k^*}.
\]
Define a function $V_3$ as
\[
V_3(\mathbf{u}(t))=\max\{Z(\mathbf{u}(t))-1,0\}+\max\{1-z(\mathbf{u}(t)),0\}.
\]

\noindent It is easily verified that $V_3$ is positive definite with respect to $\mathbf{R}^*$, i.e., (a) $V_3(\mathbf{u}(t))\geq 0$, and (b) $V_3(\mathbf{u}(t))=0$ if and only if $\mathbf{u}(t)=\mathbf{R}^{*}$. Next , let us show that $D^+V_3(\mathbf{u}(t)) \leq 0$, where $D^+$ stands for the upper-right Dini derivative. To this end, we need to show the following two claims for $t > 0$.

\emph{Claim 2:} $D^+Z(\mathbf{u}(t))\leq0$ if $Z(\mathbf{u}(t))\geq1$.
Moreover, $D^+Z(\mathbf{u}(t))<0$ if $Z(\mathbf{u}(t))>1$.

\emph{Claim 3:} $D_+z(\mathbf{u}(t))\geq0$ if $z(\mathbf{u}(t))\leq1$.
Moreover, $D_+z(\mathbf{u}(t))> 0$ if  $z(\mathbf{u}(t))<1$. Here $D_+$ stands for the lower-right Dini derivative.

\emph{Proof of Claim 2:} Choose $k_0$ such that
\[
Z(\mathbf{u}(t))=\frac{u_{k_0}(t)}{R_{k_0}^*}, \quad D^+Z(\mathbf{u}(t))=\frac{u_{k_0}^{'}(t)}{R_{k_0}^*}.
\]
Then,
\[
  \frac{R_{k_0}^{*}}{u_{k_0}(t)}u_{k_0}^{'}(t)
  =\left(1-u_{k_0}(t)\right)\frac{R_{k_0}^{*}}{u_{k_0}(t)}f_{k_0}^U(\mathbf{u}(t))-\delta_{k_0}^R R_{k_0}^{*}.
\]
If $f_{k_0}(\mathbf{u}(t))=0$, then
$\frac{R_{k_0}^{*}}{u_{k_0}(t)}u_{k_0}^{'}(t) <0$, which implies $D^+Z(\mathbf{u}(t))<0$. Now assume $f_{k_0}^U(\mathbf{u}(t))>0$, then
\[
\begin{split}
\frac{R_{k_0}^{*}}{u_{k_0}(t)}u_{k_0}^{'}(t) &\leq
(1-R_{k_0}^{*})\frac{R_{k_0}^{*}}{u_{k_0}(t)}f_{k_0}^U(\mathbf{u}(t))-\delta_{k_0}^R R_{k_0}^{*}
\leq (1-R_{k_0}^{*})f_{k_0}^U\left(\frac{R_{k_0}^{*}}{u_{k_0}(t)}\mathbf{u}(t)\right)-\delta_{k_0}^R R_{k_0}^{*} \\
& \leq (1-R_{k_0}^{*})f_{k_0}^U\left(\mathbf{R}^*\right)-\delta_{k_0}^R R_{k_0}^{*}=0,
\end{split}
\]
where the second inequality follows from the concavity of $f_{k_0}^U$, and the third inequality follows from the monotonicity of $f_{k_0}^U$. This implies $D^+Z(\mathbf{u}(t))\leq0$. Noting that the first inequality is strict if $Z(\mathbf{u}(t))>1$, we get that $D^+Z(\mathbf{u}(t))<0$ if $Z(\mathbf{u}(t))>1$. Claim 2 is proven.

The argument for Claim 3 is analogous to that for Claim 2 and hence is omitted. Next, consider three possibilities.

Case 1: $Z(\mathbf{u}(t)) < 1$. Then $z(\mathbf{u}(t)) < 1$ and $V_3(\mathbf{u}(t)) = 1 - z(\mathbf{u}(t))$. Hence,
$D^+V_3(\mathbf{u}(t)) = -D_+z(\mathbf{u}(t)) < 0$.

Case 2: $z(\mathbf{u}(t)) > 1$. Then $Z(\mathbf{u}(t)) > 1$ and $V_3(\mathbf{u}(t)) = Z(\mathbf{u}(t)) - 1$. Hence,
$D^+V_3(\mathbf{u}(t)) = D^+Z(\mathbf{u}(t)) < 0$.

Case 3 If $Z(\mathbf{u}(t)) \geq 1$, $z(\mathbf{u}(t)) \leq 1$. Then $V_3(\mathbf{u}(t)) = Z(\mathbf{u}(t)) - z(\mathbf{u}(t))$ and
$D^+V_3(\mathbf{u}(t)) = D^+Z(\mathbf{u}(t)) - D_+z(\mathbf{u}(t)) \leq 0$.
Moreover, the equality holds if and only if $\mathbf{u}(t) = \mathbf{R}^*$.

The declared result follows from the LaSalle Invariance Principle.


\bibliography{<your-bib-database>}



\end{document}